\documentclass[a4paper,12pt]{article}
\usepackage[cmex10]{amsmath}
\usepackage{latexsym,amssymb}
\usepackage[mathscr]{eucal}
\usepackage{amsxtra,amscd,amsthm,upref,layout,color}
\usepackage{calc}
\usepackage[final]{graphicx}
\def\refup#1{{$^{#1}$}}
\def\p{^{^{\prime}}}\def\pp{^{^{\prime\prime}}}

\def\ctg{\rm{ctg}}
\def\btheta{{\bar \theta}}
\def\bY{\bar{ Y}}
\def\bZ{\bar{Z}}\def\bx{\bar{x}}\def\bphi{\bar{\phi}}
\def\oo{{\leavevmode\setbox0=\hbox{h}\dimen0=\ht0 \advance\dimen0
by-1ex\rlap{\raise0.47\dimen0\hbox{\char'27}}o}}

\def\br{{\bf r}}

\def\begeq{\begin{equation}}
\def\endeq{\end{equation}}
\def\begdis{\begin{displaymath}}
\def\enddis{\end{displaymath}}

\def\cA{{\cal A}}\def\cC{{\cal C}}\def\cD{{\cal D}}

\def\cL{{\cal L}}  \def\cM{{\cal M}}
\def\cV{{\cal V}}

\def\cbDEA{{\bar\cD}_{E,A}}\def\cbDEB{{\bar\cD}_{E,B}}

\def\F12{{{_{_1}}F_{_2}}}
\def\min{\rm {min}}\def\max{\rm {max}}\def\cotg{\rm cotg}
\def\arcotg{\rm{arcotg}}\def\arctan{\rm{arctan}}\def\arcos{\rm{arcos}}\def\arcsin{\rm{arcsin}}
\def\clea{\cL_{\rm{E,1}}(r)}\def\cleb{\cL_{\rm{E,2}}(r)}
\def\clec{\cL_{\rm{E,3}}(r)}\def\cled{\cL_{\rm{E,4}}(r)}
\def\DV{{\cD}_V}
\def\DVx{{\cD}_{V,x}}\def\DVxa{{\cD}_{V,x,a}}\def\DVxb{{\cD}_{V,x,b}}
\def\DVxc{{\cD}_{V,x,c}}\def\DVxd{{\cD}_{V,x,d}}
\def\DVY{{\cD}_{V,Y}}\def\DVYa{{\cD}_{V,Y,a}}\def\DVYb{{\cD}_{V,Y,b}}\def\DVYc{{\cD}_{V,Y,c}}
\def\DVYd{{\cD}_{V,Y,d}}
\def\DVA{{\cD}_{V,A}}\def\bDVA{{\bar \cD}_{V,A}}\def\DVB{{\cD}_{V,B}}\def\DVFB{{\cD}_{V,F_B}}\def\bDVB{{\bar \cD}_{V,B}}
\def\bteara{{\bar\theta}_{\rm{E,A,1}}(r)}\def\btearb{{\bar\theta}_{\rm{E,A,2}}(r)}

\def\bphear{{\bar\phi}_{\rm{E,A}}(r,\theta)}
\def\btVR{{\bar\theta}_{V,R}}\def\btVB{{\bar\theta}_{V,B}}\def\btVx{{\bar\theta}_{V,x}}
\def\cld{\cC(r)}\def\cf{\gamma(r)}\def\dcf{\gamma\p (r)}\def\ddcf{\gamma\pp (r)}
\def\ddcfe{{\gamma_{E}}\pp (r)}\def\ddcfv{{\gamma_{ V}}\pp (r)}\def\ddcfp{{\gamma_{P}}\pp (r)}
\def\ddcfea{{\gamma_{E,A}}\pp (r)}
\def\ddcfeA{{\gamma_{E,a}}\pp (r)}\def\ddcfeB{{\gamma_{E,b}}\pp (r)}\def\ddcfeC{{\gamma_{E,c}}\pp (r)}\def\ddcfeD{{\gamma_{E,d}}\pp (r)}

\def\ddcfeb{{\gamma_{E,B}}\pp (r)}

\def\ddcfva{{\gamma_{V,A}}\pp (r)}\def\ddcfvb{{\gamma_{V,B}}\pp (r)}

\def\ddcfvA{{\gamma_{V,a}}\pp (r)}\def\ddcfvB{{\gamma_{V,b}}\pp (r)}\def\ddcfvC{{\gamma_{V,c}}\pp (r)}
\def\ddcfvD{{\gamma_{V,d}}\pp (r)}
\def\ddcfpa{{\gamma_{P,A}}\pp (r)}\def\ddcfpb{{\gamma_{P,B}}\pp (r)}

\def\ddcfpA{{\gamma_{P,a}}\pp (r)}\def\ddcfpB{{\gamma_{P,b}}\pp (r)}\def\ddcfpC{{\gamma_{P,c}}\pp (r)}\def\ddcfpD{{\gamma_{P,d}}\pp (r)}

\def\ie{{\em i.e.}}
\def\etal{{\em et al.}}
\def\hw{{\hat \omega}} \def\hnu{{\hat {\nu}}}

\begin{document}
\title{The chord-length  probability density of the regular octahedron}
\author{  
{{Salvino Ciccariello}}\\
  \begin{minipage}[t]{0.9\textwidth}
   \begin{flushleft}
   \setlength{\baselineskip}{12pt}
{\slshape  {\footnotesize{Universit\`{a} di Padova,
   Dipartimento di Fisica {\em G. Galilei}}}}\\
  {\slshape{\footnotesize{Via Marzolo 8, I-35131 Padova, Italy}} }\\
  \footnotesize{salvino.ciccariello@unipd.it}
\end{flushleft}
\end{minipage}
}
\date{ January 16, 2014\\    revised February 8, 2014}
\maketitle                        
\begin{abstract} \noindent
The chord length probability density of the regular octahedron is explicitly
evaluated  throughout its full range of distances by separating it into three
contributions respectively due to the  pairs of facets opposite to each other
or sharing an edge or a vertex.
\end{abstract}
\vfill
\eject
{}{}
\section{Introduction}
Small-angle scattering\refup{1-3} (SAS) is a powerful tool to get information on
the interface size and shape of a sample. In fact, the observed scattering intensity
is the square modulus of the Fourier transform (FT) of $n(\br)$, the scattering density
of the sample, or, equivalently, the FT of the convolution of $n(\br)$ by itself. In the
SAS realm $n(\br)$ can fairly be approximated by a discrete value function that takes
only two values in most of the cases. This approximation implies that the sample is either
a bi-continuous or a particulate system. Confining ourselves to the second case, if one
further assumes that particles have the same shape and size, that are isotropically
distributed and that their number density is small (conditions fairly met in the case of
biological samples), then the scattering density is
proportional to the FT of the isotropic correlation function of the particle defined as
\begeq\label{1.1}
\gamma(r)\equiv(1/4\,\pi\,V)\int d\hw\int \rho_V(\br_1)\rho_V(\br_1+r\hw)dv_1.
\endeq
Here $\rho_V(\br)$ denotes the characteristic function of set $\cV$ (having volume $V$)
occupied by the particle
(\ie\  it is defined as being equal to one inside the particle and to zero elsewhere).
Further, the inner integral is performed over $\cV$ (or the all space) and the outer
integral over all the directions of the unit vector $\hw$.\\
As first pointed out by Debye \etal\refup{4}, the right hand side of (\ref{1.1}),
suitably scaled,  can be interpreted  as
the probability density that by  randomly tossing a stick of length $r$ both ends of the
stick fall within the particle. This remark shows the stochastic meaning of $\cf$.
Consequently,  SAS theory is intimately related to stochastic geometry\refup{5} as
well as to integral geometry\refup{6} that aims to get general properties investigating
suitable integrals over the particle volume or surface. Interestingly, the derivatives
of $\cf$ can be expressed as integrals over the particle surface\refup{7,8} and, in the
case of convex particles, the first and second derivatives (after being appropriately scaled)
can be  interpreted as the
probability densities  for respectively finding the stick
with one end or with both ends on the particle surface. The investigations of these
integral relations yield some general result as the Porod\refup{9} and the
Kirste-Porod\refup{10} relations as well as the particle surface
features that yield discontinuous $\ddcf$s\refup{11-13}. \\

\begin{figure}[!h]
{
\includegraphics[width=7.truecm]{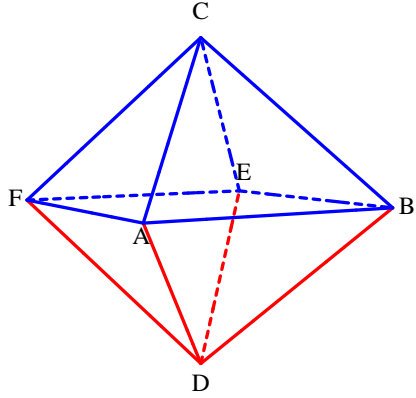}
\hskip 5.truecm
}
\caption{\label{Fig1}{{The regular octahedron.  }}}
\end{figure}
These considerations illustrate the importance of determining the explicit expressions
of $\cf$, $\dcf$ or $\ddcf$ for specific particle shapes. In fact, these expressions
were obtained in the cases of spheres\refup{2},  cubes\refup{14}, cylinders\refup{15},
right parallelepipeds\refup{16}, rotational ellipsoids\refup{17} and regular
tetrahedrons\refup{18}.\\
This paper  goes one step further since it determines  the chord-length probability
density $\cld$ of another Platonic solid: the regular octahedron
(see Fig. 1).

\section{Basic integral expression}
\noindent
In order to evaluate $\cld$ one starts from the general integral expression of
$\ddcf$ obtained by Ciccariello \etal\refup{7}, namely
\begeq\label{2.1}
\ddcf =-\frac{1}{4\pi V}\int d\hw \int_{S_1} dS_1 \int_{S_2} dS_2 (\hnu_1(\br_1)\cdot \hw)
(\hnu_2(\br_2)\cdot \hw) \delta(\br_1+r\hw -\br_2).
\endeq
Here $S1=S_2=S$ and V respectively denote the surface and volume of the octahedron,
$\hnu_1(\br_1)$ [$\hnu_2(\br_2)$] the unit normal (pointing outwardly to the octahedron)
to the infinitesimal surface element
$dS_1$ [$dS_2$] of $S$, located at the point with position vector $\br_1$ [$\br_2$].
The Dirac function $\delta(\cdot)$, present in (\ref{2.1}),
requires that the distance between $dS_1$ and $dS_2$ be equal to $r(\ge 0)$ since
$\hw$ denotes a unit vector which, according to the leftmost integral of (\ref{2.1}), ranges
over all the possible directions. Thus, at fixed $dS_1$, the integrals over $d\hw$ and
$dS_2$ amounts to integrating over the curve resulting from the
intersection of the sphere of radius $r$  and centered at $dS_1$  with $S$. \\
To explicitly evaluate expression (\ref{2.1}) it is convenient to account for the fact
that $S$ is formed by eight triangular faces.  Thus, expression (\ref{2.1}) reduces to
a sum of terms having still the form of (\ref{2.1}) but with the important changes that
integration domains $S_1$ and $S_2$ are two facets of the octahedron.
Clearly the only cases where $S_1$ and $S_2$ are different is important, otherwise the
correspondent integrand vanishes because $\hw$ is perpendicular
both to $\hnu_1$ and $\hnu_2$. Then, for each couple $(S_1,\,S_2)$,  the facets
share an edge or a vertex or, in the negative case, they lie oppositely and the
associated planes are parallel. In the last case, for conciseness, the facets will be said
parallel. For each couple $(S_1,\,S_2)$, expression (\ref{2.1}) defines a scalar
$r$-function. Hence,  whatever $S_1$ and $S_2$, expression (\ref{2.1}) takes the same form,
denoted by $\ddcfe$, if  $S_1$ and $S_2$ share an edge. The same happens if $S_1$ and $S_2$
share a vertex  or are parallel and the corresponding integrals will be denoted by $\ddcfv$
and $\ddcfp$, respectively. Thus, it results that
\begeq\label{2.2}
\ddcf=24\,\ddcfe + 24\,\ddcfv\,+8\,\ddcfp ,
\endeq
and the evaluation of $\ddcf$ reduces to that of $\ddcfe$, $\ddcfv$ and $\ddcfp$. The
evaluations will be carried out in the following three sections assuming that the
edge length $\ell$ of the octahedrons be equal to one. This assumption is by no
way restrictive. In fact, if one respectively denotes by $\gamma\pp_{\ell}(r)$ and
$\ddcf$ the functions relevant to the octahedrons with edges equal to $\ell$ and 1, one has
$\gamma\pp_{\ell}(r)=(1/\ell^2)\gamma\pp(r/\ell)$.
\section{Evaluation of $\ddcfe$}
The geometrical configuration relevant to  $\ddcfe$ is shown in Fig. 2. Surfaces $S_1$ and $S_2$
respectively are the regular triangles ABC and ABD with unit sides. It is now put
\begeq\label{3.1}
\beta = \hat{BAC}=\pi/3,\quad \alpha=\hat{DOC}=\arccos(-1/3),\quad \alpha_c=\pi-\alpha,
\endeq
\begeq\label{3.2}
H=OC=OD=\frac{\sqrt{3}}{2},\quad h=D_0D=H\sin\,\alpha_c=\sqrt{\frac{2}{3}},\  \  \  |DC|=\sqrt{2}.
\endeq
\begin{figure}[!h]
{
\includegraphics[width=7.truecm]{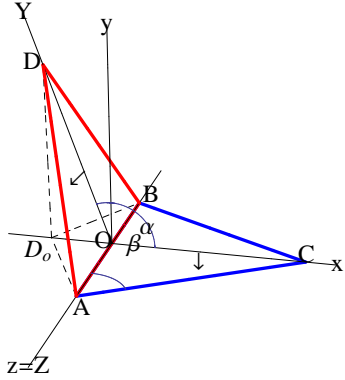}
\hskip 5.truecm
}
\caption{\label{Fig2}{{Cartesian frames used for evaluating the chord-length probability density
(CLPD) contribution due to a couple of  plane facets with a common edge. }}}
 \end{figure}\\
The minimax distances\refup{11} are
\begeq\label{3.3}
d_1=h,\quad d_2=H,\quad d_3=1\quad {\rm and} \quad d_4=\sqrt{2}.
\endeq
In terms of the axes shown in Fig. 2, one has
\begeq\label{3.4}
dS_1=dx\,dz\quad {\rm and} \quad dS_2=dY\,dZ.
\endeq
The integration domain of $dS_1$ (\ie\ the triangle ABC) is defined by the inequalities
\begeq\label{3.5}
0< x < H \quad {\rm and} \quad -L_E(x) < z <  L_E(x),
\endeq
and that of $dS_2$ by
\begeq\label{3.6}
0< Y < H \quad {\rm and} \quad -L_E(Y) < Z <  L_E(Y),
\endeq
with
\begeq\label{3.6a}
L_E(x)\equiv z_A-x\,\ctg \beta=\frac{1}{2}-x\,\ctg \beta.
\endeq
Further $\hnu_1=(0,-1,0)$ and $\hnu_2=(-\sin\alpha,\cos\alpha,0)$. With $\hw=(\hw_x,\hw_y,\hw_z)=
(\cos\varphi\,\sin\theta,\,\,\sin\varphi\,\sin\theta,\,\,\cos\theta)$ one finds that
\begeq\label{3.7}
\hnu_1\cdot\hw=  -\sin\theta\,\sin\varphi\quad{\rm and}\quad
\hnu_2\cdot\hw=-\sin\theta\,\sin(\alpha-\varphi).
\endeq
Since  $r\hw$ joins a point of ABC to a point of ABD, variable $\varphi$ varies within the interval $[\alpha,\,\pi]$. Then, putting $\varphi\equiv \varphi\p+(\alpha+\pi)/2$, it follows that
$-(\pi-\alpha)/2< \varphi\p < (\pi-\alpha)/2$ and, hereafter denoting $\varphi\p$  as $\varphi$
for notational simplicity, one gets
\begeq\label{3.8}
A_{\rm E}(\theta,\varphi)\equiv  (\hnu_1\cdot \hw)(\hnu_2\cdot\hw)\,\sin\theta = -\sin^3\theta\,\cos(\varphi+\alpha/2)\,\cos(\varphi-\alpha/2),
\endeq
In terms of the present variables the Dirac function requires that
\begin{eqnarray}&Z & =\bZ(r,z,\theta)\equiv z + r\,\cos\theta,\nonumber\\
&Y& =\bY(r,\theta,\varphi)\equiv r\,\cos(\varphi+\alpha/2)\sin(\theta)/\sin\alpha,\label{3.9}\\
&x& =\bx(r,\theta,\varphi)\equiv \bY(r,\theta,\varphi)\cos\alpha+ r\,\sin(\varphi+\alpha/2)\,\sin\theta=\bY(r,\theta,-\varphi).\nonumber
\end{eqnarray}
The $x$, $Y$ and $Z$ integrals are immediately performed and, taking into account that
the Jacobian related to the $Y$ variable is $1/\sin \alpha$,   one finds that
\begeq\label{3.10}
\ddcfe=-\frac{1}{4\pi V\sin\alpha}\int_0^{\pi}d\theta \int_{-\frac{\pi-\alpha}{2}}^{\frac{\pi-\alpha}{2}}
A_{E}(\theta,\varphi)\Theta_{ E}(\bx,\bY)d\varphi\int_{-L_E(\bx)}^{L_E(\bx)}\Theta_z(\bZ)dz,
\endeq
where function $\Theta_{E}(\bx,\bY)$, depending on  $\theta$ and $\varphi$,
is equal to one  if  $\bx$ and  $\bY$ obey inequalities (\ref{3.5}a) and (\ref{3.6}a)
and to zero elsewhere. Similarly $\Theta_z(\bZ)$ is equal to one or zero depending on
whether $\bZ$ obeys inequalities (\ref{3.6}b) or not. According to (\ref{3.9}a) $\bZ$
linearly depends on $z$. Then, the above $z$-integral is equal to
\begin{align}
F_{E}(r,\theta,\varphi)& \equiv \min[L_E(\bx),\,L_E(\bY)-r\cos\theta]-\max[-L_E(\bx),\, -L_E(\bY)-r\cos\theta]=\nonumber \\
&\min[L_E(\bx),\,L_E(\bY)-r\cos\theta]+\min[L_E(\bx),\, L_E(\bY)+r\cos\theta]\label{3.11}
\end{align}
if the above quantity is positive otherwise it is equal to zero. It is now observed that
each of the following transformations $\theta\, \to\, (\pi-\theta)$ and $\varphi\, \to\,-\varphi$
interchanges $\bx$ with $\bY$ and leaves expressions (\ref{3.8}) and (\ref{3.11}) invariant.
Thus,
(\ref{3.10}) becomes
\begeq\label{3.12}
\ddcfe=-\frac{1}{\pi V\sin\alpha}\int_{0}^{\pi/2} d\theta \int_{0}^{(\pi-\alpha)/{2}}
A_{\rm E}(\theta,\varphi)F_{E}(r,\theta,\varphi)d\varphi,
\endeq
with the  constraints
\begeq\label{3.13}
0<\,\bx(r,\theta,\varphi)\,< H,\quad 0<\,\bY(r,\theta,\varphi)\,< H, \quad
F_{\rm E}(r,\theta,\varphi)\,>\,0
\endeq
that will generally reduce the integration domain
\begeq\label{3.14}
\cD_E\equiv\{0< \theta < \pi/2,\  \
0<\varphi<(\pi-\alpha)/2\}
\endeq
to a smaller one. All the constraints
must explicitly be reduced in order to get the explicit expression of $\ddcfe$.
To this aim we start from the constraints implicit in definition (\ref{3.11}). Putting
\begeq\label{3.15}
\mu_{\pm}(\theta,\varphi)\equiv 2\ctg \beta\, \sin\theta\, \cos(\varphi\pm \alpha/2)/sin\alpha
\endeq
one easily shows that $F_{\rm E}(r,\theta,\varphi)$ becomes
\begeq\label{3.16}
F_{\rm E}(r,\theta,\varphi)=1-\frac{r}{2}\,\bigl(\max[\mu_{+}+\cos\theta,\,\,\mu_{-}-\cos\theta]+
\max[\mu_{+}-\cos\theta,\,\, \mu_{-}+\cos\theta]\bigr).
\endeq
The condition $(\mu_{+}+\cos\theta)\,>\,(\mu_{-}-\cos\theta)$ is fulfilled within the $\cD_E$ sub-domain
\begeq\label{3.17}
\cD_{E,A}\equiv \{\,0<\varphi<(\pi-\alpha)/2,\ \ \ 0\,<\, \theta\, <\,\btheta_{E}(\varphi)\equiv \arcotg(\sin\varphi)\},
\endeq
and $(\mu_{-}+\cos\theta)\,>\,(\mu_{+}-\cos\theta)$ throughout $\cD_E$.
Then, from equations (\ref{3.16}) and (\ref{3.1}) follows that $F_{\rm E}(r,\theta,\varphi)$  takes the
forms
\begeq\label{3.18}
F_{\rm {E,A}}(r,\theta,\varphi)\equiv\, 1 -r\bigl(\cos\theta+\frac{\sin\theta\,\cos\varphi}{\sqrt{2}}\bigr)\quad {\rm if}\quad
(\theta,\varphi)\in \cD_{E,A},
\endeq
 and
\begeq\label{3.19}
F_{\rm {E,B}}(r,\theta,\varphi)\equiv 1 -r\, \sqrt{3/2}\,\,\sin\theta\,\cos(\varphi-\alpha/2)
\quad {\rm if}\quad
(\theta,\varphi)\in \cD_{E,B},
\endeq
$\cD_{E,B}$ being defined as the complementary domain of $\cD_{E,A}$ in $\cD_E$.
\subsection{Constraints and integration domains}
Before proceeding to the integral evaluation one must still
reduce  inequalities (\ref{3.13}a), (\ref{3.13}b) and (\ref{3.13}c). The  last requires
that $F_{\rm {E,A}}>0$ within $\cD_{E,A}$ and that
$F_{\rm {E,B}}>0$ within $\cD_{E,B}$. The inequalities depend on $r$ and the same will
happen for the associated domains that will be determined confining ourselves to $\cD_E$
and to the non-trivial $r$-domain $[0,\,\sqrt{2}]$ of $\ddcf$.  \\
For inequalities (\ref{3.13}a) one finds that the associated domain $\cD_{E,x}(r)$ is
\begin{eqnarray}
\cD_{E,x}(r)  &= & \cD_E \quad\quad\quad\quad\quad\quad\quad\ {\rm if}\quad\quad 0 < r < \sqrt{3}/2, \label{3.20}\\
\cD_{E,x}(r)&=&\cD_{E,x,1}(r)\cup \cD_{E,x,2}(r)\quad {\rm if}\quad \sqrt{3}/2 < r < \sqrt{2},\label{3.21}
\end{eqnarray}
with
\begin{eqnarray}
\cD_{E,x,1}(r)&\equiv&\{\,0<\theta<\pi/2,\ 0\,<\varphi\,<\bphi_{E,x}(r) \},\label{3.22}\\
\cD_{E,x,2}(r)&\equiv&\{\,\bphi_{E,x}(r)\,<\varphi\,<(\pi-\alpha)/2,\ 0<\theta<\btheta_{E,x}(r,\varphi)\},\label{3.23}
\end{eqnarray}
and
\begin{eqnarray}
\bphi_{E,x}(r)&\equiv& \alpha/2 - \arcos(\sqrt{2/3}\big/r),\label{3.24}\\
\btheta_{E,x}(r,\varphi)&\equiv& \arcsin\bigl(\sqrt{2/3}\big/[r\cos(\varphi-\alpha/2)]\bigr).\label{3.25}
\end{eqnarray}
Inequalities (\ref{3.13}b) are fulfilled throughout $\cD_E$ if \ $0<r<\sqrt{2}$. \\
Inequality $F_{\rm {E,B}}(r,\theta,\varphi)>0$ coincides with $\bx(r,\theta,\varphi)\,< H$ and
is therefore obeyed throughout $\cD_{E,F_B}\equiv\cD_{E,x}$. \\
The reduction of the last inequality $F_{\rm {E,A}}(r,\theta,\varphi)>0$ yields the domain
$\cD_{E,F_A}(r)$ defined as
\begin{eqnarray}
\cD_{E,F_A,a}(r)  &\equiv & \cD_{E,A} \quad\quad\quad\quad\quad\quad\quad\quad
 \quad {\rm if}\quad\quad 0 < r < \sqrt{2/3}, \label{3.26}
 \end{eqnarray}
 to
 \begin{equation}\label{3.27}
\cD_{E,F_A,b}(r)=
\begin{cases}
\cD_{E,A,1}(r)  \equiv \{ 0<\theta<\clea ,\  \  \  0<\varphi<(\pi-\alpha)/2\}, \\
\cD_{E,A,2}(r) \equiv \{ \clea <\theta<\cleb,\ \  \bphear<\varphi<(\pi-\alpha)/2\} \\
\cD_{E,A,3}(r)  \equiv \{ \cleb <\theta<\pi/2,\ \ \ 0<\varphi<(\pi-\alpha)/2\},
\end{cases}
\end{equation}
if $\sqrt{2/3}\, <\, r \,<\,\sqrt{3}\big/2$, to
 \begin{equation}\label{3.28}
\cD_{E,F_A,c}(r)=
\begin{cases}
\cD_{E,A,1}(r)  , \\
{\cD\p}_{E,A,2}(r) \equiv \{ \clea <\theta<\clec,\  \  \
\bphear<\varphi<(\pi-\alpha)/2\}, \\
{\cD\p}_{E,A,3}(r)  \equiv \{ \cled <\theta<\cleb, \  \  \  \bphear<\varphi<(\pi-\alpha)/2\},\\
{\cD\p}_{E,A,4}(r)  \equiv \{\cleb <\theta<\pi/2,\  \  \  0<\varphi<(\pi-\alpha)/2\}
\end{cases}
\end{equation}
if $\sqrt{3}\big/2\,<\,r\,<\,1$,   and to
\begin{equation}\label{3.29}
\cD_{E,F_A,d}(r)=
\begin{cases}
{\cD\pp}_{E,A,1}(r) \equiv  \{ \cled<\theta<\cleb,\ \ \ \bphear<\varphi<(\pi-\alpha)/2\}, \\
{\cD\pp}_{E,A,2}(r)  \equiv \{\cleb  <\theta<\pi/2,\ \ \ 0<\varphi<(\pi-\alpha)/2\}
\end{cases}
\end{equation}
if $1\,<\,r\,<\,\sqrt{2}$.
\begin{figure}[!h]
\includegraphics[width=7.truecm]{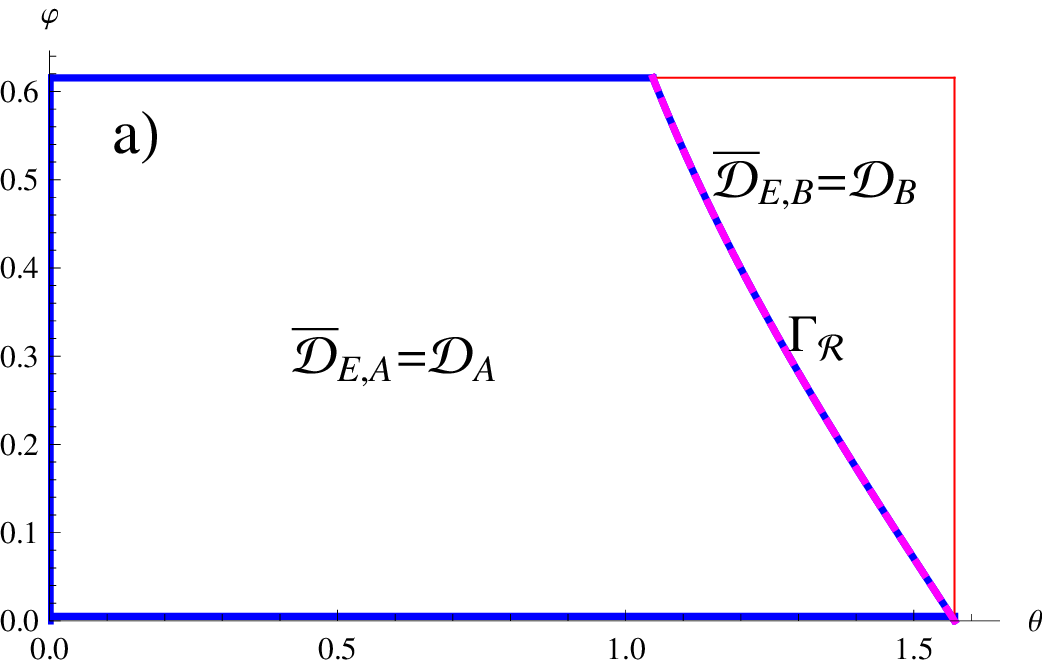}\includegraphics[width=7.truecm]{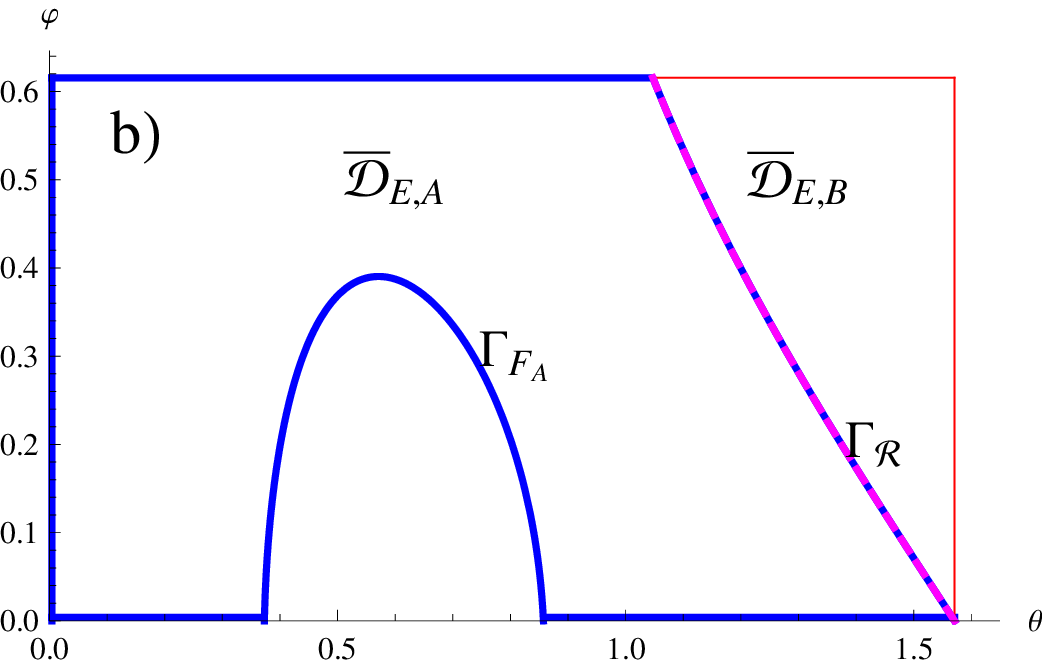}
\includegraphics[width=7.truecm]{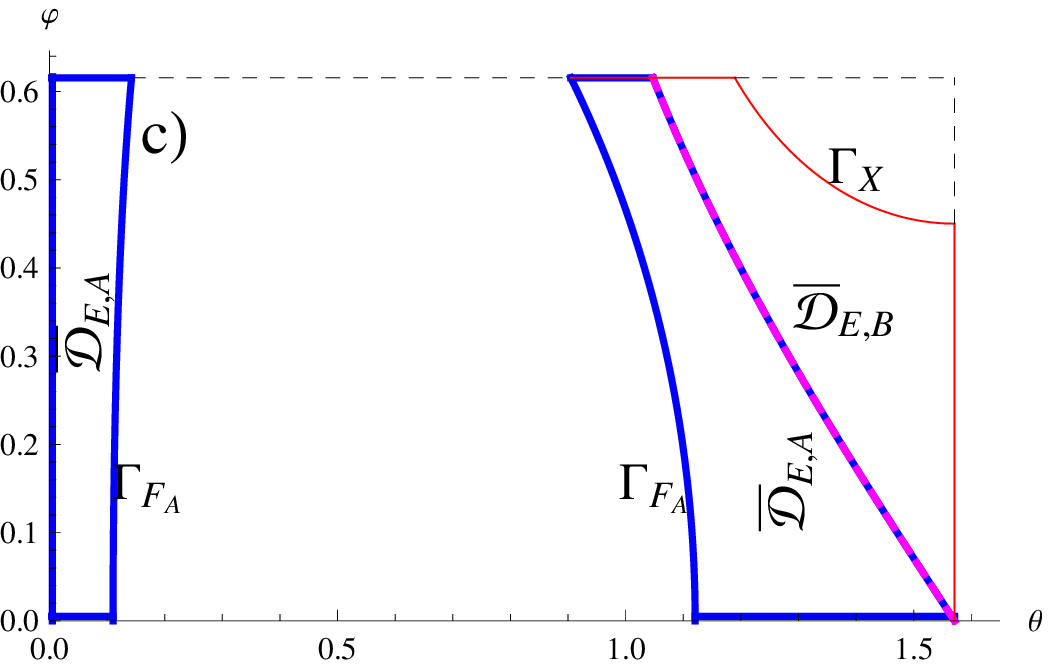}\includegraphics[width=7.truecm]{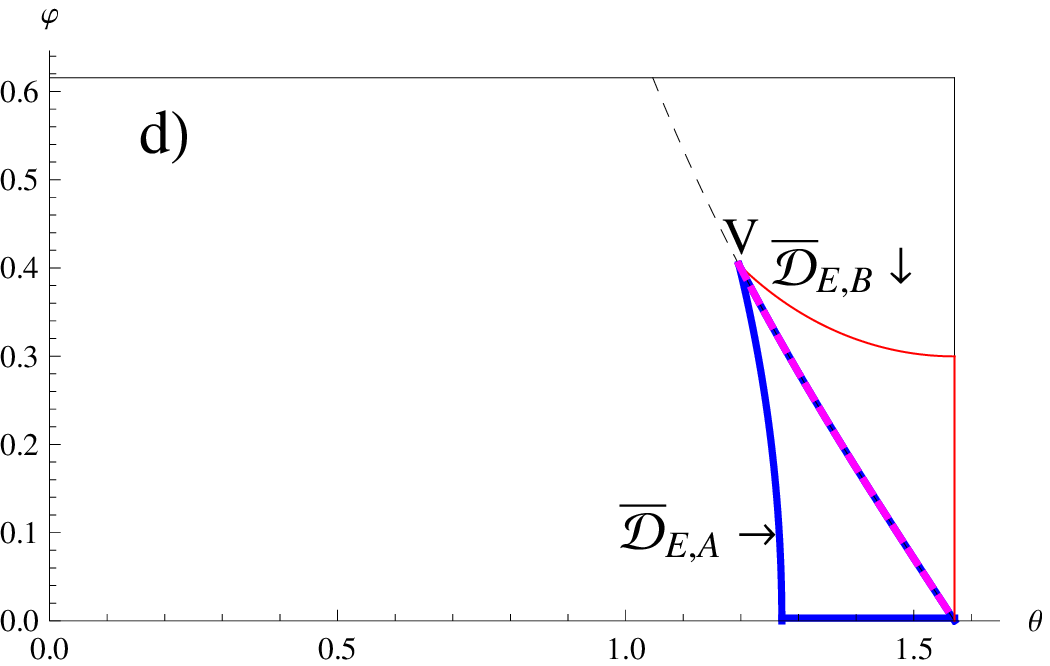}
\caption{\label{Fig3}{{The four panels shows the  shapes of the integration domains $\cbDEA(r)$
and $\cbDEB(r)$ in the four $r$-ranges: $a=[0,\ \sqrt{2/3}]$ ,  $b=[\sqrt{2/3},\ \sqrt{3}\big/2]$,
$c=[\sqrt{3}\big/2,\ 1]$ and $d=[1,\ \sqrt{2}]$. The $\cbDEA(r)$ domains are bounded by thick blue
lines and the $\cbDEB(r)$s by red continuous ones. The dotted thick curve $\Gamma_R$ separates
domain $\cD_{E,F_A}$ from  $\cD_{E,F_B}$ and is defined by Eq. (\ref{3.17}). The bell shaped curve
$\Gamma_{F_A}$ is related to the condition $F_{E,A}>0$ [see Eq.s (\ref{3.27})-(\ref{3.29})] and
curve $\Gamma_x$ to the boundary of $\cD_{E,x}$ [see Eq.s (\ref{3.20})-(\ref{3.25})].  }}
}
\end{figure}
The involved $\cL_{..}(r)$ functions are defined as follows
\begeq\label{3.38a}
\clea\equiv \bteara-\alpha/2, \quad  \cleb\equiv \pi - \alpha/2 -\bteara
\endeq
\begeq\label{3.38b}
\clec\equiv \btearb-\pi/3, \quad  \cled\equiv 2\pi/3-\btearb
\endeq
with
\begeq\label{3.39}
\bteara\equiv \arcsin(\sqrt{2/3}\big/r),
\endeq
\begeq\label{3.40}
\btearb\equiv \arcsin(\sqrt{3}\big/2r),
\endeq
\begeq\label{3.42}
\bphear\equiv\arccos\bigl(\sqrt{2}(1-r\cos\theta)\big/(r\sin\theta)\bigr).
\endeq
\subsection{$\ddcfe$ expression}
The  $\ddcfe$ expression is given by the sum of the following two expressions
\begin{eqnarray}
\ddcfea&\equiv&-\frac{1}{\pi V\sin\alpha}\int_{\cbDEA(r)}
A_{\rm E}(\theta,\varphi)F_{\rm {E,A}}(r,\theta,\varphi)d\theta d\varphi\label{3.42a}\\
\ddcfeb&\equiv&-\frac{1}{\pi V\sin\alpha}\int_{\cbDEB(r)}
A_{\rm E}(\theta,\varphi)F_{\rm {E,B}}(r,\theta,\varphi)d\theta d\varphi\label{3.43}
\end{eqnarray}
with
\begin{eqnarray}
\cbDEA(r)&\equiv&\cD_{E,F_A}\cap\cD_{E,x}(r)\cap\cD_{E,A}(r)\label{3.44}\\
\cbDEB(r)&\equiv&\cD_{E,B}\cap\cD_{E,x}(r).\label{3.45}
\end{eqnarray}
The shapes of these domains are shown in Fig. \ref{Fig3} in the four ranges of distances
{\em a, b, c}\  and $d$, defined in the caption. The evaluation of the integrals is a long
task that was made possible by the MATHEMATICA software. By these results
the final $\ddcfe$ expression is
\begin{eqnarray}
\ddcfeA&\equiv& \frac{2 \sqrt{2} - \pi  + \alpha}{8 \pi }
 -\frac{(18 + (7 \sqrt{3}-9 ) \pi )\,r }{96 \sqrt{2}\, \pi } ,\label{3.46} \\
\ddcfeB &\equiv& \frac{1}{12\, \sqrt{6}\, r^3} - \frac{1}{2\, \sqrt{6}\, r}  +
\frac{4 + \sqrt{2}\,(\pi  + \alpha )}{8\, \sqrt{2}\, \pi }- 
\frac{(18 - (9 - 13 \sqrt{3}) \pi)\,  r}{96 \sqrt{2}\,\, \pi},\label{3.47}\\
\ddcfeC&\equiv&  -\frac{1}{288\, \sqrt{2}\, \pi\,  r^3} \bigg[-4 \sqrt{3}\,\pi  - 36 (4 + \sqrt{2}\, \alpha )\, r^3 + 3 (18 - 9 \pi  + 10 \sqrt{3}\, \pi )\, r^4  +\nonumber \\
 &&6\, (25 r^2-6)\, R_{34}(r)+8 \sqrt{3}\, \arcsin\Big[\frac{9 r^2-7 }{2 R_{23}^3(r)}\Big] +\nonumber\\
 &&   96\, \sqrt{3}\, r^2\, \arcsin\Big[\frac{1}{2 R_{23}(r)}\Big] -
   72 \sqrt{2}\, r^3\, \arcsin\Big[\frac{r}{\sqrt{3}\, R_{23}(r)}\Big] -\label{3.48} \\
 &&   18 \sqrt{3}\,r^4\, \arcsin\Big[\frac{\sqrt{3}\, (27 - 90 r^2 + 96 r^4 - 34 r^6 + 2 r^8)}
   {2\, r^7\, R_{23}(r)}\Big]\bigg]\nonumber,\\
\ddcfeD&\equiv& -\frac{1}{192 \pi\,  r^3}\bigg[ -16 \sqrt{2} +
   8 \sqrt{2}\, (3 + \sqrt{3}\, \pi )\, r^2 - 12 \pi \, r^3 +\nonumber \\
 &&   3 \,\sqrt{2}\, (4 \sqrt{3}-3)\, \pi \, r^4 -
   4\, \sqrt{2}\, (5 r^2-2)\, R_{11}(r) -\nonumber \\
 &&   24\, r^3\, \arcsin\Big[\frac{4 + 4 r^2 - 7 r^4}{R_{23}^4(r)}\Big] +
   36\, \sqrt{2}\,\, r^4 \arcsin\Big[\frac{R_{11}(r)}{r}\Big] - \nonumber \\
 &&   8 \sqrt{6}\, r^2\, (2 + 3 r^2)\, \arcsin\Big[\frac{1 + 3 R_{11}(r)}{2 R_{23}(r)}\Big]\bigg].\label{3.49}
\end{eqnarray}
Here suffices $a,\, b,\, c$ and $d$ specify the $r$-range where the expressions apply
and the following definitions  have also been used
\begeq\label{3.54}
R_{11}(r) = \sqrt{r^2 - 1},\quad R_{23}(r) = \sqrt{3\,r^2-2},\quad
R_{34}(r) = \sqrt{4\,r^2-3}.
\endeq 
\section{Evaluation of $\ddcfv$}
\begin{figure}[!h]
{\includegraphics[width=7.truecm]{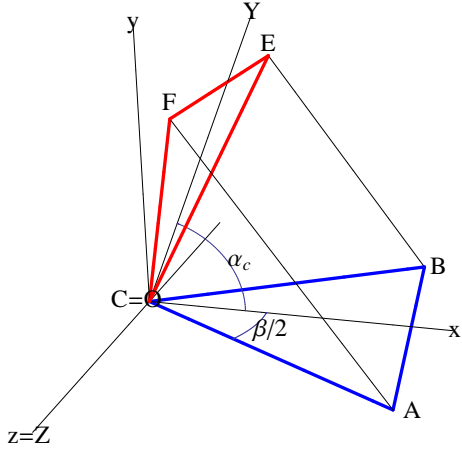}
\hskip 5.truecm }
\caption{\label{Fig4}{{Cartesian frames used for evaluating the CLPD contribution due to a
couple of  plane facets sharing a vertex. }}}
\end{figure}

One passes now to evaluate the CLPD relevant to a pair of facets sharing a vertex. The
configuration is shown in Fig. \ref{Fig4} that also shows the Cartesian frames used to
work out the CLPD expression. The evaluation proceeds along the same route described in
section 3. The integration domains $S_1=ABC$ and $S_2=CEF$ are defined by the inequalities
\begeq\label{4.1}
0 < x < H, \quad -L_V(x)\, <z\,<\, \, L_V(x)
\endeq
and
\begeq\label{4.2}
0 < Y < H, \quad -L_V(Y)\, <\,Z\,<\, \, L_V(Y)
\endeq
with
\begeq\label{4.3}
L_V(x)\equiv x\tan(\beta/2)=x/\sqrt{3}.
\endeq
The unit normals are $\nu_1=(0,-1,0)$ and $\nu_2=(-\sin\alpha_c,\cos\alpha_c,0)$ with $\alpha_c\equiv(\pi-\alpha)$. After putting $\theta=\theta'+\pi/2$  and
$\varphi=\varphi'+(\pi+\alpha_c)/2$ (and denoting $\theta'$ and $\varphi'$ again by $\theta$
and $\varphi$ for notational simplicity) one finds that $|\theta|<\pi/2$ and
$|\varphi|<(\pi-\alpha_c)/2=\alpha/2$. The angular factor $(\nu_1\cdot\hw)
(\nu_2\cdot\hw)\sin\theta$ becomes
\begeq\label{4.4}
\cA_V(\theta,\varphi)=-\cos^3\theta\cos(\varphi+\alpha_c/2)\cos(\varphi-\alpha_c/2).
\endeq
The Dirac function and the linearity of $L_V(x)$ makes the evaluation of the $x$, $z$, $Y$
and $Z$ integrals straightforward and $\ddcfv$ takes the form
\begeq\label{4.5}
\ddcfv=-\frac{1}{\pi V}\int_0^{\pi/2}d\theta\int_0^{\alpha/2}
\cA_V(\theta,\varphi)F_V(r,\theta,\varphi)d\varphi
\endeq
with
\begin{align}\label{4.6}
F_V(r,\theta,\varphi) \equiv & r\bigl(\min[a(\theta,\varphi)+(\sin\theta)/2,\, b(\theta,\varphi)-(\sin\theta)/2]+\nonumber \\
&\quad\quad \min[a(\theta,\varphi)-(\sin\theta)/2,\, b(\theta,\varphi)+(\sin\theta)/2]\bigr)
\end{align}
and
\begin{eqnarray}
a(\theta,\varphi)&\equiv&\cotg\beta\,\cos\theta\,\cos(\varphi+\alpha_c/2)\big/\sin\alpha_c,\label{4.7}\\
b(\theta,\varphi)&\equiv&\cotg\beta\,\cos\theta\,\cos(\varphi-\alpha_c/2)\big/\sin\alpha_c.\label{4.8}
\end{eqnarray}
The integrand of (\ref{4.5}) is invariant with respect to each of the following two transformations
$\theta\to -\theta$ and $\varphi\to -\varphi$. This explains the integration bounds reported in Eq.
(\ref{4.5}) and the omission of factor 4 at the denominator. Further, $F_V(r,\theta,\varphi)$ must
be positive and this condition will generally make the integration domain smaller than
$\DV\equiv\{0<\theta<\pi/2,\, 0<\varphi<\alpha/2\}$.
Solving the inequalities implicit in (\ref{4.6}) one finds that $F_V(r,\theta,\varphi)$ is equal to
\begeq\label{4.8a}
F_{V,A}(r,\theta,\varphi) = 2r[\cotg\beta\,\cos\theta\cos(\varphi+\alpha_c/2)]\big/\sin\alpha_c
\endeq
within the integration sub-domain
\begeq\label{4.9}
\cD_{V,A} \equiv\{0<\theta<\btVR(\varphi), \ 0<\varphi<\alpha/2\}
\endeq
and to
\begeq\label{4.10}
F_{V,B}(r,\theta,\varphi) = r[\cotg\beta\,\cos\theta\cos\varphi-\sin\theta\,\sin(\alpha_c/2)]\bigr/sin(\alpha_c/2)
\endeq
within the integration sub-domain
\begeq\label{4.11}
\cD_{V,B}  \equiv\{\btVR(\varphi)<\theta<\pi/2, \ 0<\varphi<\alpha/2\}
\endeq
with
\begeq\label{4.12}
\btVR(\varphi)\equiv \arctan[\sin(\varphi)/\sqrt{2}].
\endeq
Both $F_{V,A}(r,\theta,\varphi)$ and $F_{V,B}(r,\theta,\varphi)$ must be non negative. One
easily verifies that $F_{V,A}(r,\theta,\varphi)>0$ within $\cD_{V,F_A}=\DVA$ and that $F_{V,B}(r,\theta,\varphi)>0$ within the subset $\DVFB$ of $\DVB$ defined as
\begeq\label{4.13}
\DVFB\equiv \{ \btVR(\varphi)<\theta<\btVB(\varphi), \ 0<\varphi<\alpha/2\}
\endeq
with
\begeq\label{4.14}
\btVB(\varphi)\equiv \arctan(\cos\varphi).
\endeq
\subsection{Constraints and integration domains for $\ddcfv$}
To determine the integration domains of $F_{V,A}(r,\theta,\varphi)$  and
$F_{V,B}(r,\theta,\varphi)$ one must also require that the $x$ and $Y$ values,
determined by the vanishing of the arguments of the Dirac function and equal to
\begeq\label{4.15}
{\bar x}_{V}(r,\theta,\varphi)\equiv r\cos\theta\cos(\varphi-\alpha_c/2)\,/\sin\alpha_c
\endeq and
\begeq\label{4.16}
{\bar Y}_{V}(r,\theta,\varphi)\equiv r\cos\theta\cos(\varphi+\alpha_c/2)\,/\sin\alpha_c,
\endeq
respectively, obey constraints (\ref{4.1}) and (\ref{4.2}), i.e.
\begeq\label{4.17}
0\,<\,{\bar x}_{V}(r,\theta,\varphi)\,<\,H\quad{\rm and}\quad
0\,<\,{\bar Y}_{V}(r,\theta,\varphi)\,<\,H.
\endeq
\begin{figure}[!h]
{\includegraphics[width=7.truecm]{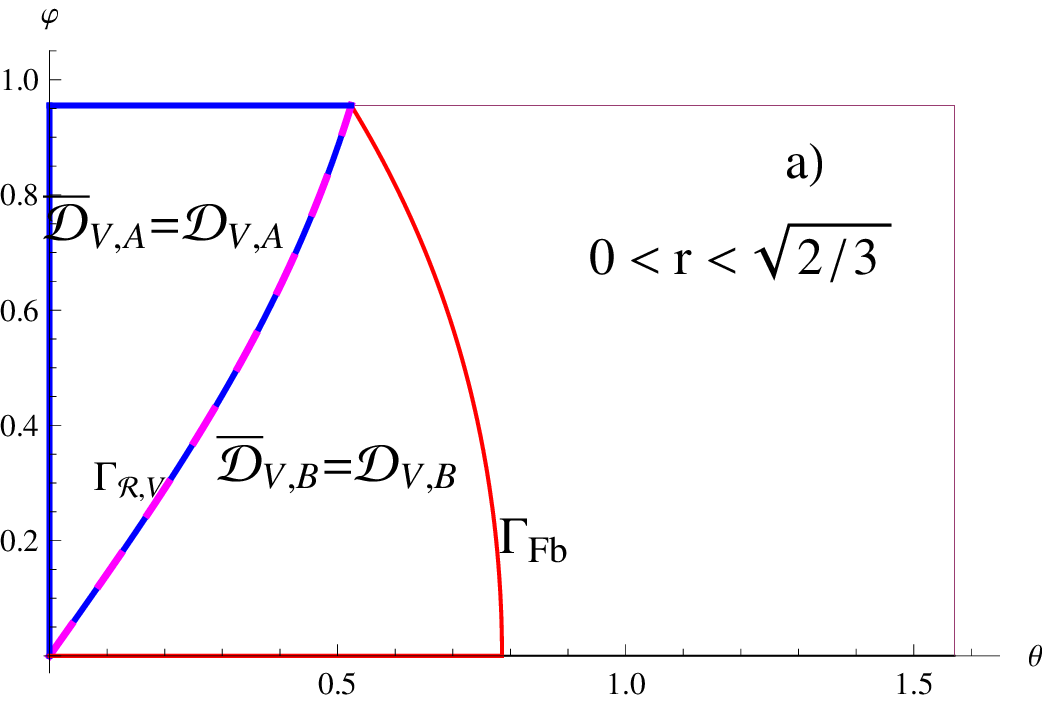}\includegraphics[width=7.truecm]{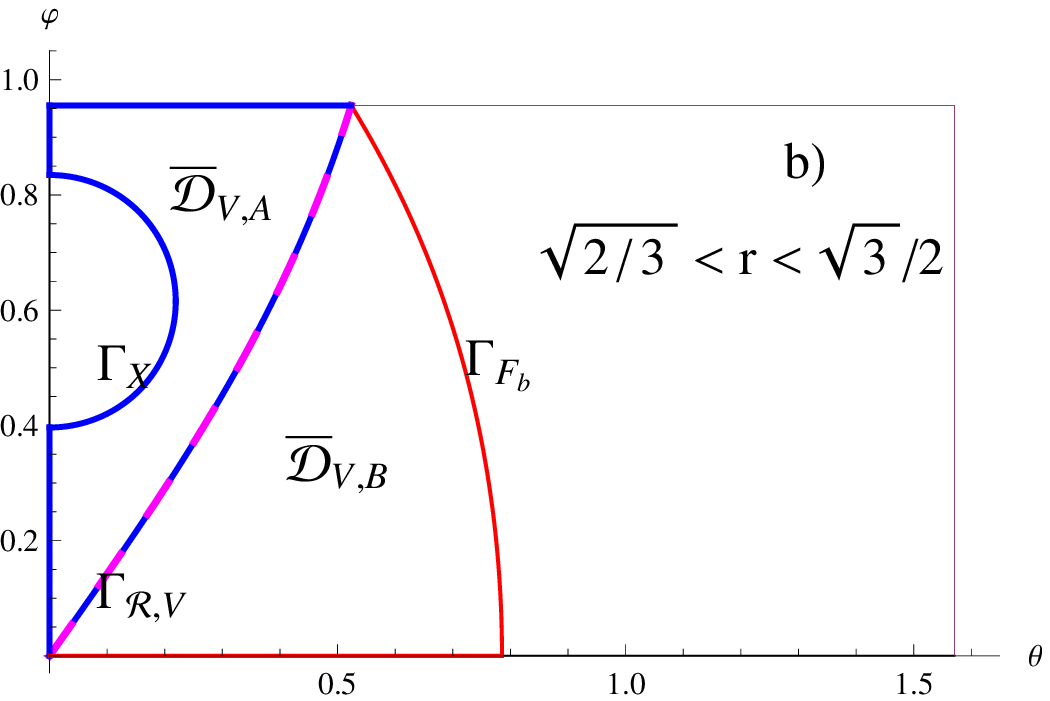}
\includegraphics[width=7.truecm]{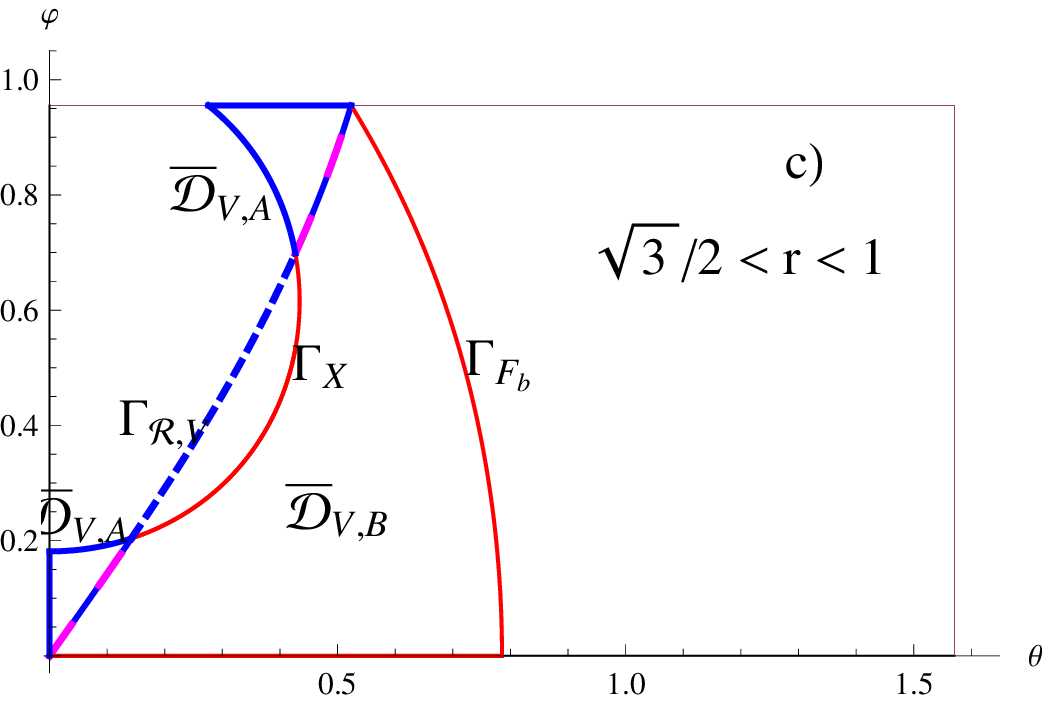}\includegraphics[width=7.truecm]{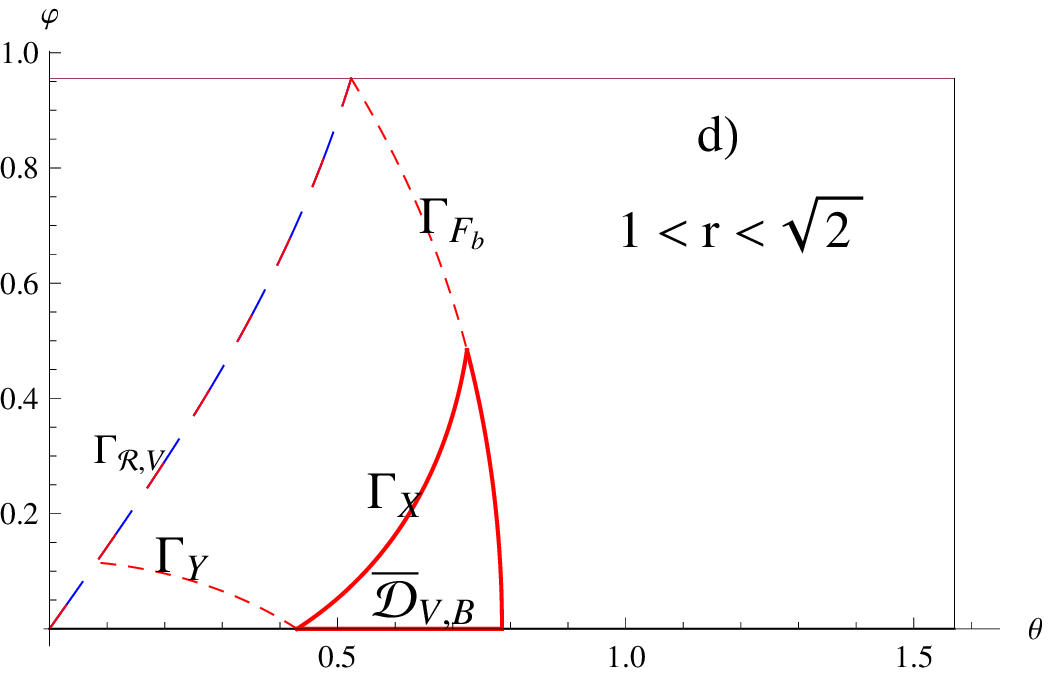}
\hskip 5.truecm }
\caption{\label{Fig5}{{Four panels a), b), c) and d)  show the typical shapes of the integration
domains $\bDVA(r)$ and $\DVFB(r)$ in the four $r$ intervals  $a,\,\,b,\,\,c$[ and $d$.
The $\bDVA(r)$ domains are bounded by thick blue lines and the $\bDVB(r)$s by red ones. The
dotted thick curve separates $\cD_{V,A}$ from $\cD_{V,B}$.  Curves $\Gamma_{V,R}$,
$\Gamma_{V,x}$, $\Gamma_{V,Y}$ and $\Gamma_{F_B}$ are respectively
defined by Eq.s  (\ref{4.12}), (\ref{4.18}),  (\ref{4.27}) and  (\ref{4.14}). }}}
\end{figure}
The domain $\DVx(r)$, where inequalities (\ref{4.17}a) are fulfilled, is determined by
curve $\Gamma_{V,x}(r)$, defined (where existing) as
\begeq\label{4.18}
\Gamma_{V,x}(r)\equiv \{\btVx (r,\varphi)\equiv \arcos[\sqrt{2/3}\big/r\cos(\varphi-\alpha_c/2)],
\, \varphi\},
\endeq
and explicitly reads in the four $r$-ranges $a,\,\,b,\,\,c$ and $d$
\begeq\label{4.19}
\DVxa(r)=\DV,
\endeq
\begin{equation}\label{4.20}
\DVxb(r)=
\begin{cases}
\{0<\theta<\pi/2,\ 0<\varphi<{\bar\varphi}_{V,x,-}(r)\}, \\
\{{\bar\theta}_{V,x}(r,\varphi)<\theta<\pi/2,\
{\bar\varphi}_{V,x,-}(r)<\varphi<{\bar\varphi}_{V,x,+}(r)\},    \\
\{0<\theta<\pi/2,\  {\bar\varphi}_{V,x,+}(r) < \varphi<\alpha/2\},
\end{cases}
\end{equation}
\begin{equation}\label{4.21}
\DVxc(r)=
\begin{cases}
\{0<\theta<\pi/2,\ 0<\varphi<{\bar\varphi}_{V,x,-}(r)\}, \\
\{{\bar\theta}_{V,x}(r,\varphi)<\theta<\pi/2,\
{\bar\varphi}_{V,x,-}(r)<\varphi<\alpha/2\},
\end{cases}
\end{equation}
\begeq\label{4.22}
\DVxd(r)=\{\btVx(r,\varphi)<\theta<\pi/2,\ 0<\varphi<\alpha/2\},
\endeq
with
\begeq\label{4.23}
{\bar\varphi}_{V,x,\pm}(r)\equiv \alpha_c/2\pm\arcos(\sqrt{2/3}\bigl/r).
\endeq
The $\DVY(r)$ domain,  where inequalities (\ref{4.3}b) are obeyed, in the four distance
sub-ranges is
\begin{eqnarray}
\DVYa(r)&=&\DV,\label{4.24}\\
\DVYb(r)&=&\DV,\label{4.24b}\\
\DVYc(r)&=&\DV,\label{4.24c}\\
\DVYd(r)&=&\begin{cases}\{{\bar\theta}_{V,Y}(r,\varphi)<\theta<\pi/2,\ 0<\varphi<{\bar\phi}_{V,Y}(r)\},\\
\{0\,<\,\theta\,<\pi/2,\,\,{\bar\phi}_{V,Y}(r)\,<\,\varphi\,<\alpha/2\}
\end{cases}\label{4.24d}
\end{eqnarray}
with                
\begin{eqnarray}
{\bar\theta}_{V,Y}(r,\varphi)&\equiv& \arcos[\sqrt{2/3}/(r\cos(\varphi+\alpha_c/2))],\label{4.25}\\
{\bar\phi}_{V,Y}(r)&\equiv&-\,{\bar\varphi}_{V,x,-}(r).\label{4.26}
\end{eqnarray}
\subsection{$\ddcfv$ expression}
Similarly to subsection 3.2, the  $\ddcfv$ expression is now given by the sum of the following two expressions
\begin{eqnarray}
\ddcfva&\equiv&-\frac{1}{\pi V\sin\alpha_c}\int_{\bDVA(r)}
A_{ V}(\theta,\varphi)F_{{V,A}}(r,\theta,\varphi)d\theta d\varphi\label{4.27}\\
\ddcfvb&\equiv&-\frac{1}{\pi V\sin\alpha_c}\int_{\bDVB(r)}
A_{V}(\theta,\varphi)F_{ {V,B}}(r,\theta,\varphi)d\theta d\varphi\label{4.28}
\end{eqnarray}
with
\begin{eqnarray}
\bDVA(r)&\equiv&\cD_{V,F_A}\cap\DVx(r)\cap\DVY(r)\label{4.29}\\
\bDVB(r)&\equiv&\DVFB\cap\DVx(r)\cap\DVY(r).\label{4.30}
\end{eqnarray}
The shapes of these domains are shown in Fig. \ref{Fig5} in the four $r$-ranges
$a,\,\,b,\,\,c$ and $d$, defined in the caption of Fig. \ref{Fig5}. The integrals
were evaluated by the MATHEMATICA.  From these follows that the $\ddcfv$ expression
in the four $r$-ranges is
\begin{eqnarray}
\ddcfvA&\equiv&(9 + \sqrt{3})\, r\big/(96 \sqrt{2}),\label{4.31}\\
\ddcfvB&\equiv&-1\big/(4 \sqrt{6}\, r^3) + 1\big/(\sqrt{6}\, r) +
(9 - 29 \sqrt{3})\, r\big/(96 \sqrt{2}),\label{4.32}\\
\ddcfvC&\equiv&\frac{1}{192 \sqrt{2}\, \pi\,  r^3}\Bigg[-8 \sqrt{3} \pi  +
   4\, r^2\, \bigl(10\, \sqrt{3}\, \pi  + 3\, R_{34}(r) - 4\, \sqrt{3}\, \Lambda_c(r)\bigr) -\nonumber \\
&&   9\, (7 \sqrt{3}-2)\, \pi\,  r^4 +
   16\, \sqrt{3}\, (4\,r^2-1 )\arcsin\Big[\frac{7 - 9 r^2}{2\, R_{23}^3(r)}\Big]+ \nonumber\\
&&   2 \sqrt{3}\, r^4 \bigg[18\, \arcsin\Big[\frac{\sqrt{3}}{2 r}\Big] + 8 \arcsin\Big[\frac{2\,r^2-3}{2\,r^2}\Big] +
\arcsin\Big[\frac{9 - 12 r^2 + 2 r^4}{2\, r^4}\Big] -\nonumber \\
&&      30\, \arcsin\Big[\frac{7 - 9 r^2}{2 R_{23}^3(r)}\Big] -
      24\, \arcsin\Big[\frac{6 r^2-5}{2 R_{23}^2(r)}\Big]\bigg]\Bigg],\label{4.33}\\
\ddcfvD&\equiv& -\frac{1}{48\, \sqrt{2}\, \pi\,  r^3\, \bigl(3\, r^2 - 2\, R_{11}(r)\bigr)^2}
\bigg[3\, \big(-16 + 8 (12 + \sqrt{3}\, \pi )\, r^2 - \nonumber\\
&& 16\, (5 + \sqrt{3}\,\pi )\, r^4 -
       2\, (18 + 5 \sqrt{3}\,\pi )\, r^6 +9\, (3 + 2 \sqrt{3}\,\pi )\, r^8\big) - \nonumber \\
 &&  6\, \big(8 + 12\, r^2 - 2\, (31 + 6 \sqrt{3}\, \pi )\, r^4 +
       3\, (9 + 4 \sqrt{3}\, \pi )\, r^6\big)\, R_{11}(r) +\label{4.35}\\
 &&   2\, r^2\, \Big(-4 + 9\, r^4 +r^2\,\big (4 - 12\, R_{11}(r)\big)\Big)\,
 \Big [-2 \sqrt{3}\, \arcsin\big(\frac{4 - 3\, r^2}{R_{23}^2(r)}\big) +\nonumber\\
 &&      3 r^2\,\Big(3\, \arcsin\big(\frac{R_{11}(r)-1}{\sqrt{2}\, r}\big) +
          2\, \sqrt{3}\, \arcsin\big(\frac{1}{R_{23}(r)}\big)\Big) -
       2\, \sqrt{3}\, R_{23}^2(r) \Lambda_D(r)\Big]\bigg],\nonumber
\end{eqnarray}   
with  the following definitions
\begin{equation}\label{4.36}
\Lambda_C(r)\equiv
\begin{cases}
  \arcsin\big(\frac{17 + 18\, r^2\, (-2 + r^2)}{2\, R_{23}^4(r)}\big)\  &\text{if}\quad
  \sqrt{3}\,\big/2\,<\, r\, <\, \sqrt{5/6},\\
  -\pi - \arcsin\big(\frac{17 + 18\, r^2\, (-2 + r^2)}{2\, R_{23}^4(r)}\big)\ &\text{if}\quad
  \sqrt{5/6}\,<\,r\,<\,1,\\
 \end{cases}
\end{equation}   
and
\begin{equation}\label{4.37}
\Lambda_D(r)\equiv\begin{cases}
 \arcsin\Big(\frac{\sqrt{3}\, (R_{11}(r)+1)}{2 R_{23}(r)}\Big)\ &\text{if}\quad
1\,<\,r\,<\,\sqrt{10}\big/3\\
 \pi  - \arcsin\Big(\frac{\sqrt{3}\, (R_{11}(r)+1)}{2 R_{23}(r)}\Big)\ &\text{if}\quad
\sqrt{10}\big/3\,<\,r\,<\,\sqrt{2}.\\
\end{cases}
\end{equation}
\section{Evaluation of $\ddcfp$}
\begin{figure}[!h]
{ \includegraphics[width=7.truecm]{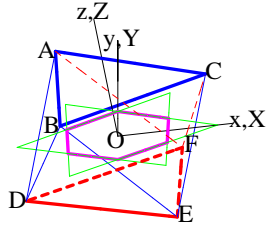}
\hskip 5.truecm }
\caption{\label{Fig6}{{Cartesian frames used for evaluating the CLPD contribution of a
couple of  plane facets sharing a vertex. The regular hexagon (of side 1/3) is the orthogonal
projection on the plane $z=0$ of the facets' portions that are each other parallel.}}}
\end{figure}

The last case of  two parallel facets (see Fig. \ref{Fig6}) is now tackled.
With the Cartesian frames shown in the figure, facets $S_1=DEF$ and $S_2=ABC$ are defined
by the following inequalities
\begeq\label{5.1}
x_m\equiv-2H/3\,<\,x\,<\,H/3\,\equiv x_M,\quad -L_P(x)\,<\,y\,<\,L_P(x)
\endeq
\begeq\label{5.2}
X_m\equiv-H/3\,<\,X\,<\,2H/3\,\equiv X_M,\quad -L_P(-X)\,<\,Y\,<\,L_P(-X)
\endeq
with
\begeq\label{5.3}
L_P(x) \equiv (x+2H/3)\tan(\beta/2).
\endeq
The unit normals are $\nu_1=(0,0,-1)$ and $\nu_2=(0,0,1)$ so that the angular factor is
\begeq\label{5.4}
\cA_P(\theta)\equiv(\nu_1\cdot\hw)(\nu_2\cdot\hw)=-\cos^2\theta.
\endeq  
The integrand of $\ddcfp$ is $\bigl(\sin\theta\, \cA_P(\theta)\delta(\br_1+r\hw-\br_2)\bigr)$ with
$\br_1=(x,y,z)$ and $\br_2=(X,Y,Z)$. The reflection with respect to the plane $y=0$ leaves
the configuration shown in Fig.\ref{Fig6} invariant. Thus, the integrand is left invariant by the transformation $\varphi\to(2\pi-\varphi)$. Hence the $\varphi$ integral can be restricted to the
interval $[0,\,\pi]$ provided the result be multiplied by two. The Dirac function fixes the
values of $\theta$, $\varphi$ and $Y$ as follows
\begeq\label{5.5}
\cos{\bar \theta}=h/r,\quad \cos{\bar \varphi}=(X-x)/(r\sin{\bar \theta})
\endeq
and
\begeq\label{5.6}
{\bar Y}=y+r\,\sin{\bar \theta}\,\sin{\bar \varphi}.
\endeq
The integral over $\theta$, $\varphi$ and $Y$ yields   
\begeq\label{5.7}
\ddcfp=-\int_{x_m}^{x_M} d x\int_{X_m}^{X_M} d X \int_{-L_P(x)}^{L_P(x)}d y
\frac{\cA_P({\bar \theta})\Theta_{{\bar P}}\Theta_{{\bar \varphi}}  }
{2\pi V r^2\,\sin{\bar \theta}\,\sin{\bar \varphi} },
\endeq
where
\begin{align}
\Theta_{{\bar P}}&\equiv\Theta\bigl(L_P(-X)-{\bar Y}\bigr)
\Theta\bigl(L_P(-X)+{\bar Y}\bigr),\label{5.8a}\\
\Theta_{{\bar \varphi}}&\equiv\Theta(1-|\cos{\bar \varphi}|),\label{5.8}
\end{align}  
The explicit evaluation of the $y$ integral gives
\begeq\label{5.9}
\ddcfp=\frac{h^2}{2\pi V r^3} \int_{x_m}^{x_M}d x\int_{X_m}^{X_M}d X
\frac{\Theta_{{\bar \varphi}}F_P(r,x,X) } { \Delta(r,x,X) }
\endeq
with
\begin{align}
\Delta(r,x,X)&\equiv \sqrt{r^2-h^2-(x-X)^2},\label{5.10a}\\
\cM(x,X,\Delta)&\equiv\min[L_P(x)+\Delta/2,\, L_P(-X)-\Delta/2],\label{5.10}
\end{align}
and
\begeq\label{5.11}
F_P(r,x,X)\equiv \cM(x,X,\Delta)+\cM(x,X,-\Delta)
\endeq
It is convenient to consider the new integration variables $(t,\, u)$, obtained by a rotation of
$-\pi/4$ of  $(x,X)$, i.e.
\begeq\label{5.12}
x=(t+u)\sqrt{2},\quad\quad X=(t-u)/\sqrt{2}.
\endeq
The new integration bounds are
\begeq\label{5.13}
t_m\equiv -\sqrt{3/8\,}\, <\, t\,<\,t_M\,\equiv\,\sqrt{3/8\,}
\endeq
\begeq\label{5.14}
u_m(t)\,<\,u\,<\,u_M(t)
\endeq
with
\begin{equation}\label{5.15}
\ \begin{cases}
u_m(t)\equiv t-\sqrt{2/3}& \text{if}\ \  t>0, \\
u_m(t)= u_m(-t)&\text{if}\ \  t<0,\\
u_M(t)\equiv 1\big/\sqrt{6}-t& \text{if}\ \  t>0, \\
u_M(t)= u_M(-t)&\text{if}\ \  t<0.
\end{cases}
\end{equation}
They are invariant with respect to the exchange $t\to\, -t$ which also implies that
$x\leftrightarrow -X$.
From definitions (\ref{5.10a}), (\ref{5.5}b), (\ref{5.8}), (\ref{5.10}) and (\ref{5.11}) follows that $\Delta$, $\Theta_{{\bar \varphi}}$ and $F_P(r,x,X)$ are left
invariant by the reflection $t\to\, -t$.  Then (\ref{5.9}) becomes
\begeq\label{5.16}
\ddcfp=\frac{h^2}{\pi V r^3} \int_{0}^{t_M}d t\int_{u_m(t)}^{u_M(t)}d u
\frac{{\tilde\Theta}_{{\bar \varphi}}{\tilde F}_P(r,t,u) } {{\tilde \Delta}(r,u) },
\endeq
where the function symbols with the tilde denote the results of the variable change (\ref{5.12})
on the corresponding functions without the tilde.
The reduction of the inequalities implicit in the  ${\tilde F}_P(r,t,u)$ definition leads
to the following two expressions of the integrand (leaving aside the factor $h^2/\pi V r^3$):
\begeq\label{5.17}
F_{P,A}(r,t,u)\equiv 2\bigl[{1+\sqrt{3/2\,}\,(u-t)}\bigr]\big/3{\tilde \Delta}(r,u)
\endeq
within the domain
\begeq\label{5.18}
\cD_{P,A}(r)\equiv\{t>\sqrt{3/2}\,\,{\tilde \Delta}(r,u),\ -\sqrt{(r^2-h^2)/2}\,<u< \sqrt{(r^2-h^2)/2}\}
\endeq
and
\begeq\label{5.18a}
F_{P,B}(r,t,u)\equiv \bigl[2({1+\sqrt{3/2\,}\,\,u)}/3 - {\tilde \Delta}(r,u)\bigr]\big/{\tilde \Delta}(r,u)
\endeq
within the domain
\begeq\label{5.19}
\cD_{P,B}(r)\equiv\{0<t<\sqrt{3/2}\,\,{\tilde \Delta}(r,u),\ -\sqrt{(r^2-h^2)/2}\,<u<\sqrt{(r^2-h^2)/2}\}.
\endeq
Both $\cD_{P,A}(r)$ and $\cD_{P,B}(r)$ are subsets  of the integration domain $\cD_V$ reported
in (\ref{5.16}).
\begin{figure}[!h]
{ \includegraphics[width=7.truecm]{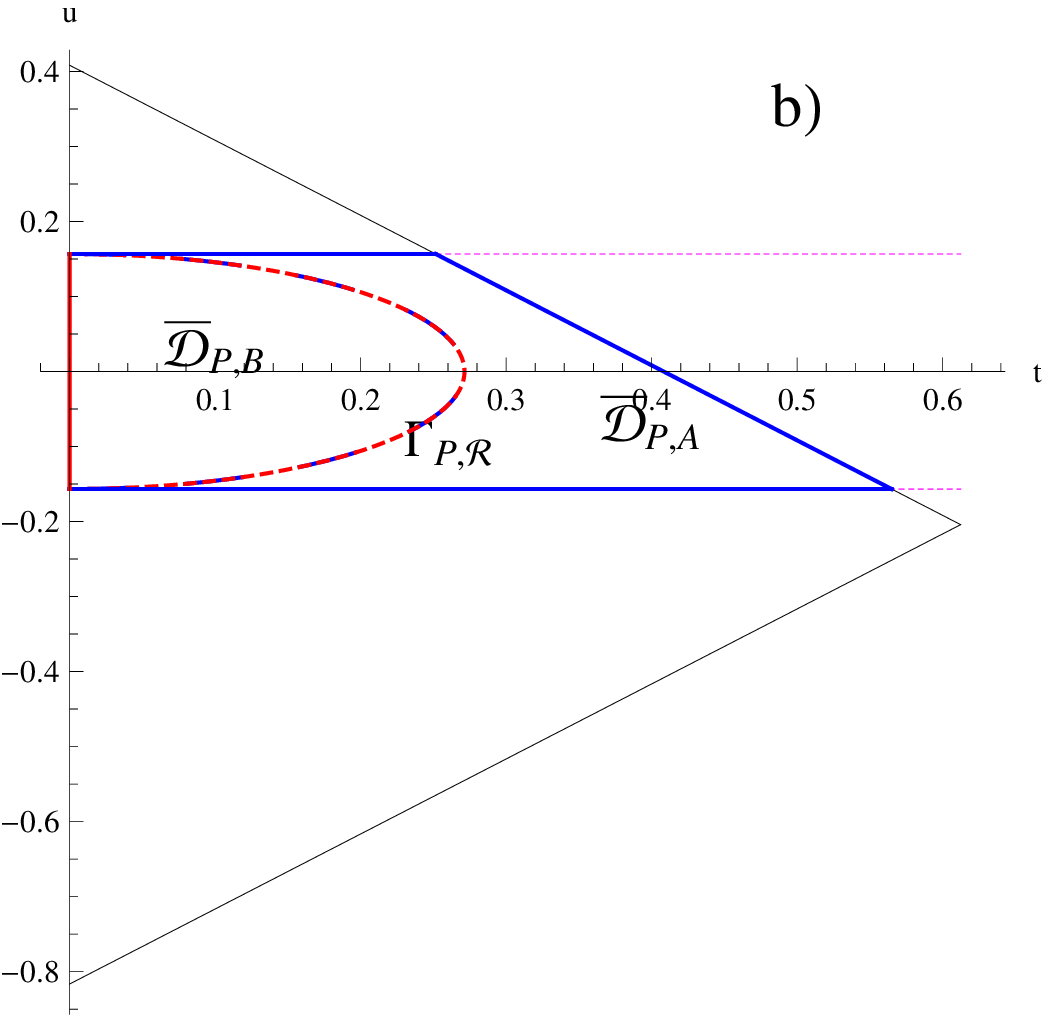}
\includegraphics[width=7.truecm]{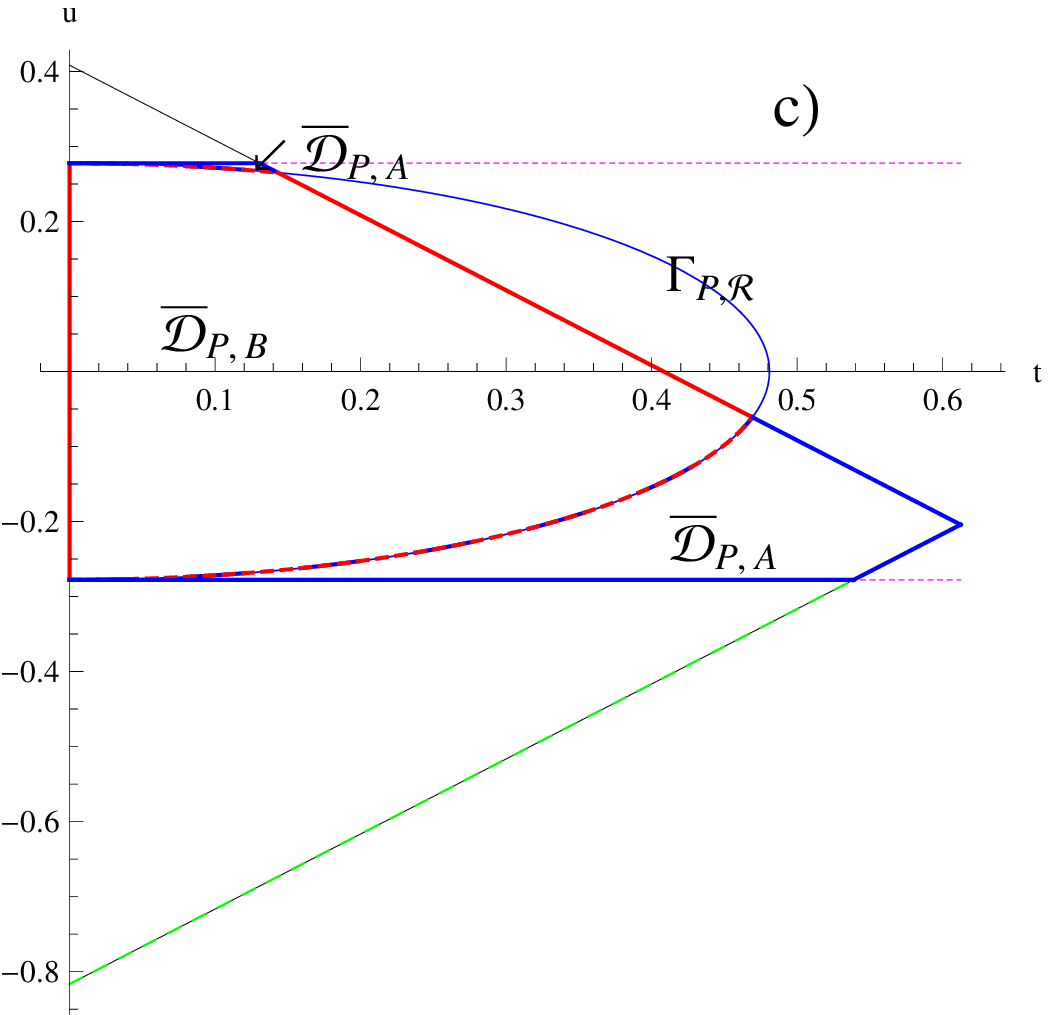}
\includegraphics[width=7.truecm]{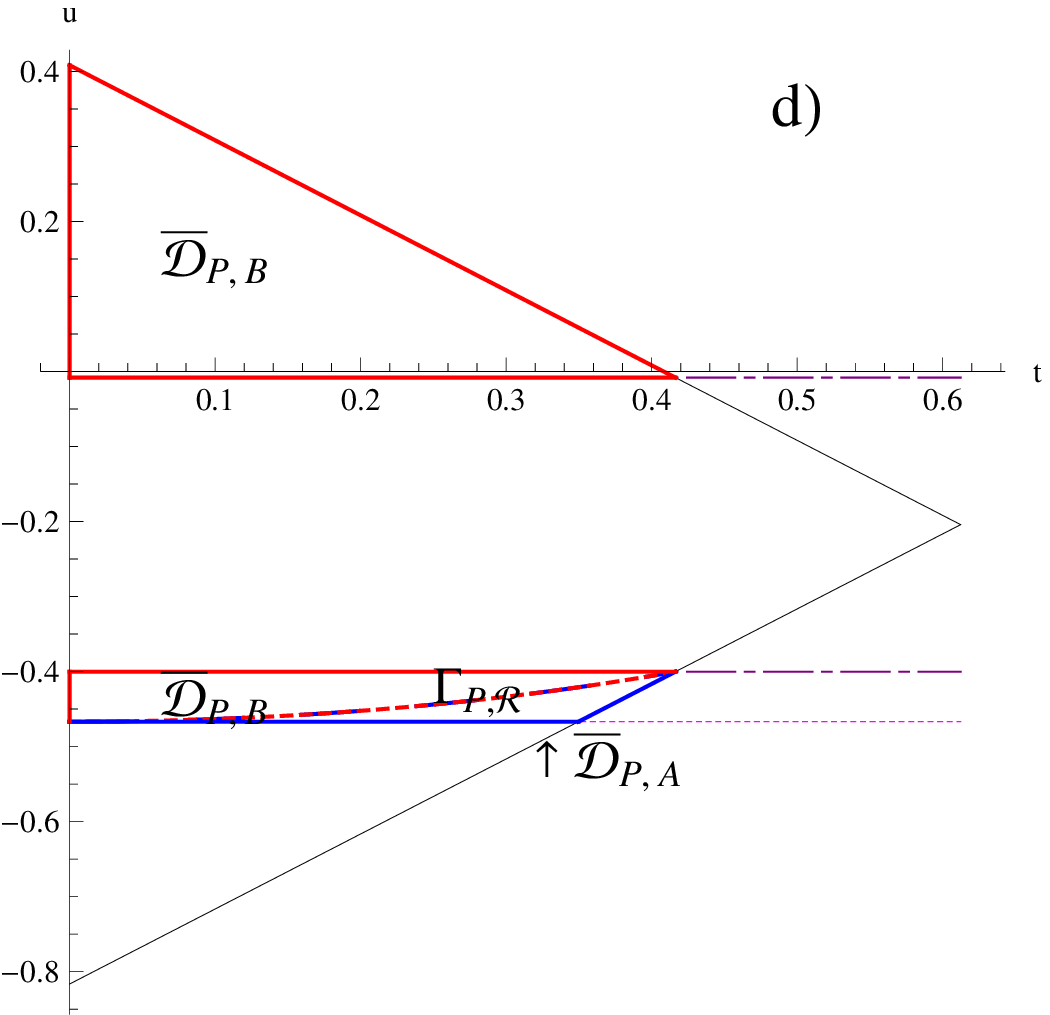}
\hskip 5.truecm   }
\caption{\label{Fig7}{{The three panels show the typical shapes of the integration domains
${\bar\cD}_{P,A}(r)$ and ${\bar\cD}_{P,B}(r)$ in the $r$ sub-ranges $b,\,\,c$ and $d$.
The ${\bar\cD}_{P,A}$ domains are bounded by thick blue lines and the ${\bar\cD}_{P,A}$s by red ones.
The dash-dot thick curve ($\Gamma_{P,R}$) separates the $\cD_{P,A}$ from $\cD_{P,B}$ [see Eq. (\ref{5.18})].
The magenta horizontal dotted
lines defines $\cD_{P,{\bar\varphi}}$ [see Eq. (\ref{5.24})]. In case $d$, the upper line is not visible because it
lies outside $\cD_P$, while the thin dash-dot purple lines arise from Eq. (\ref{5.23}).)}}}
\end{figure}
The above two integrands must be positively valued. Thus one finds that $F_{P,A}(r,t,u)>0$
within the domain
\begeq\label{5.19a}
\cD_{P,F_A}\equiv \{t>0,\ u\,>\,(t-\sqrt{2/3\,})  \}
\endeq
and that $F_{P,B}(r,t,u)>0$ within the domain $\cD_{P,F_B}(r)$ that in four $r$ sub-ranges reads
\begeq\label{5.20}
\cD_{P,F_B,a}(r)= \emptyset,
\endeq
\begeq\label{5.21}
\cD_{P,F_B,b}(r)= \{t>0,\ -\sqrt{(r^2-h^2)/2}\, < u < \,\sqrt{(r^2-h^2)/2}\},
\endeq
\begeq\label{5.22}
\cD_{P,F_B,c}(r)=\{t>0,\ -\sqrt{(r^2-h^2)/2}\, < u < \,\sqrt{(r^2-h^2)/2}\},
\endeq
\begin{equation}\label{5.23}
\cD_{P,F_B,d}(r)=
\begin{cases}
\{t>0,\   -\sqrt{(r^2-h^2)/2}\, <\,u\, <\, -(1+3\sqrt{r^2-1})\,\big/2\sqrt{6}\}, \\
\{t>0,\   (-1+3\sqrt{r^2-1})\,\big/2\sqrt{6}\, <\,u\, <\, \sqrt{(r^2-h^2)/2}\}.
\end{cases}
\end{equation}
The condition $|\cos{\bar \varphi}|<1$ is fulfilled within
\begeq\label{5.24}
\cD_{P,{\bar \varphi}}(r)=\{t>0,\ -\sqrt{(r^2-h^2)/2}\, <\, u\, < \,\sqrt{(r^2-h^2)/2}\}.
\endeq
Finally the condition $|\cos{\bar \theta}|<1$  requires that
the listed domains depending on $r$ be identified with the void one  if $r<h$.\\
In conclusion the final  integration domain  of $F_{P,A}(r,t,u)$ is
\begin{equation}\label{5.25}
{\bar\cD}_{P,A}(r)=
\begin{cases}
\emptyset &\text{if}\quad 0<r<h, \\
\cD_{P}\cap\cD_{P,A}(r)\cap\cD_{P,F_A}(r)\cap\cD_{P,{\bar \varphi}}(r)&\text{if}\quad h<r<\sqrt{2},
\end{cases}
\end{equation}
and that of $F_{P,B}(r,t,u)$ is
\begin{equation}\label{5.25a}
{\bar\cD}_{P,B}(r)=
\begin{cases}
\emptyset &\text{if}\quad 0<r<h, \\
\cD_{P}\cap\cD_{P,B}(r)\cap\cD_{P,F_B}(r)\cap\cD_{P,{\bar \varphi}}(r)&\text{if}\quad h<r<\sqrt{2}.
\end{cases}
\end{equation}
The typical shapes of these domains are illustrated in Fig. \ref{Fig7}.
\subsection{$\ddcfp$ expression}
Similarly to subsections 3.2 and 4.2, the  $\ddcfp$ expression is now given by the sum of the following two expressions
\begin{eqnarray}
\ddcfpa&\equiv&-\frac{h^2}{\pi V r^2}\int_{{\bar\cD}_{P,A}(r)}
F_{{P,A}}(r,t,u)d t du\label{5.26}\\
\ddcfpb&\equiv&-\frac{h^2}{\pi V r^2}\int_{{\bar\cD}_{P,B}(r)}
F_{P,B}(r,t,u)d t d u.\label{5.27}
\end{eqnarray}
The evaluation of the integrals  by the MATHEMATICA yields the $\ddcfpa$ and $\ddcfpb$ expressions
that in turns determine $\ddcfp$ in the four $r$-ranges as
\begin{eqnarray}
\ddcfpA &\equiv & 0, \label{5.28}\\
\ddcfpB&\equiv&\sqrt{3}\,(1 - r^2)\big/(2\, \sqrt{2}\, r^3 ),\label{5.29}\\
\ddcfpC&\equiv&\frac{1}{4 \sqrt{2}\, \pi\,  r^3}\Big[\sqrt{3} \pi - 3\, R_{34}(r) -\label{5.30}  \\
 &&  2\, \sqrt{3}\, (2 r^2-1 ) \arcsin\big(\frac{1}{2\, R_{23}(r)}\big)\Big],\nonumber\\
\ddcfpD&\equiv&-\frac{1}{36 \sqrt{2}\, \pi\,  r^3}\Big[6\, \sqrt{3}\,\pi  + (27 -
11\, \sqrt{3}\, \pi)\,  r^2 -\nonumber\\
 &&  54 R_{11}(r) + 6\, \sqrt{3}\, (5 r^2-6) \arcsin\big(\frac{1}{R_{23}(r)}\big) +\label{5.31}\\
&&   12\, \sqrt{3}\, r^2\, \arcsin\big(\frac{1 + 3 R_{11}(r)}{2 R_{23}(r)}\big)\Big].\nonumber
\end{eqnarray}         
\section{$\ddcf$ properties }
\begin{figure}[!h]
\includegraphics[width=7.truecm]{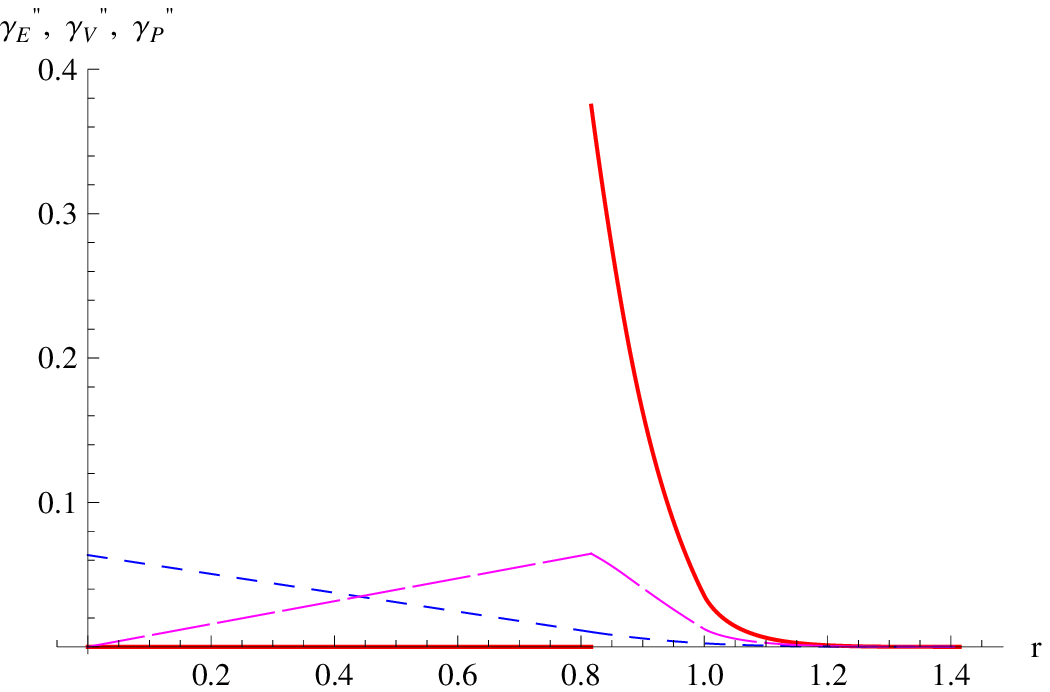}\includegraphics[width=7.truecm]{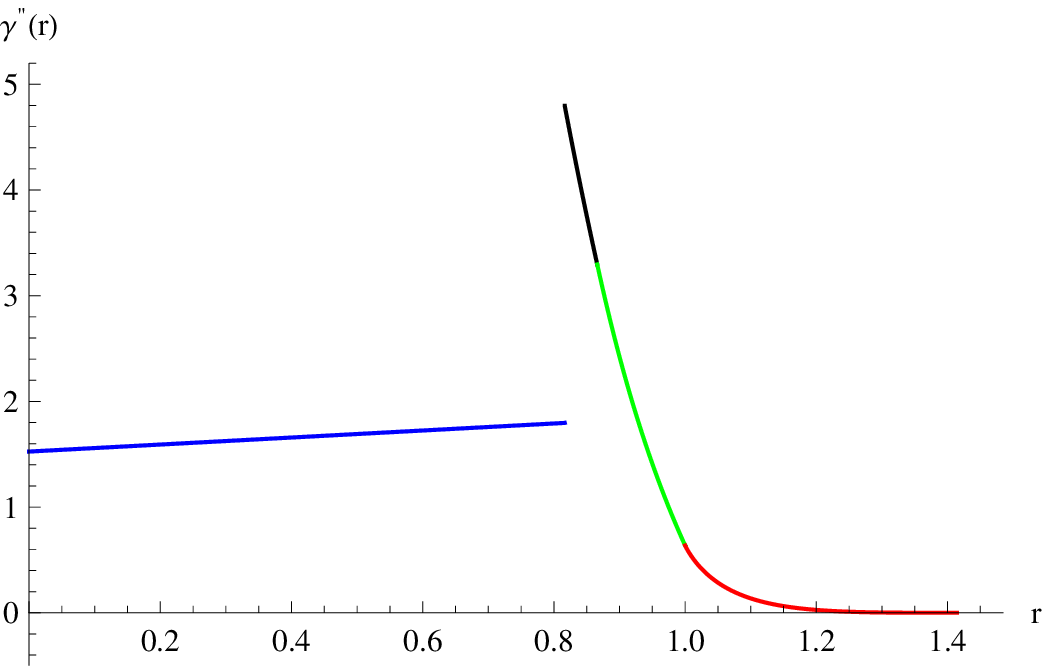}
\caption{\label{Fig8}{{In the left panel the dotted blue curve is the plot of $\ddcfe$,
the long dashed magenta curve that of $\ddcfv$ and the thick red curve that of $\ddcfp$.
The right panel shows the plot of $\ddcf$ with different colors in the four r-ranges. }}}
\end{figure}

The expression of $\ddcf$ immediately follows from (\ref{2.2}) and  the reported
expressions of $\ddcfe$, $\ddcfv$ and $\ddcfp$. The expression is not reported because one does not
have a significant cancelation of the addends. The left panel of Fig.\ref{Fig8} shows
the behaviour of $\ddcfe$, $\ddcfv$ and $\ddcfp$ and the right panel that of  $\ddcf$.\\
All the functions behave linearly in the innermost $r$-range. This result is not surprising
because it was since long proved in Ref. [19] that $\ddcf$ is a linear $r$-function
in the innermost $r$-range  whatever the particle shape provided its boundary is made up of plane facets. The two coefficients of the linear relation are related to the angularity and to
the roundness of the particle surface. Further, the explicit expression of the angularity in terms of
the edge lengths and the relevant dihedral angles is given by Eq. (4.4) of Ref. [7] and  that of
the roundness in terms of the dihedral and edge angles by Eq.s (1.7), (3.6), (3.7), (3.14) and (3.11) of Ref. [19]. One easily verifies that the linear coefficients that determine $\ddcf$ in the $r$-range {\it a} coincide with the quoted expressions.\\
Function $\ddcf$ shows a finite discontinuity at $r=\sqrt{2/3}$ that originates from the discontinuity
present in $\ddcfp$. The discontinuity is due to the fact that parts of the opposite facets of the octahedron are each other parallel. As shown in Ref.s [12,13], the presence of a parallelism, at a relative distance $d$, between subsets of the particle surface is responsible for a discontinuity in $\ddcf$. In the case of plane parallel surfaces of area  $S_p$ and distant $d$, according to Ref. [12], the discontinuity value is
\begeq\label{6.1}
\gamma\pp(d^+)-\gamma\pp(d^-)=S_p\big /(2 d V)
\endeq
In the octahedron case, $S_p$ is the area of the hexagon of side $1/3$ (see the caption of Fig.
\ref{Fig6}) and $d=h$. Since $\gamma\pp(d^+)-\gamma\pp(d^-)=8(\gamma\pp_{P,b}(h)-\gamma\pp_{P,a}(h)$,
by Eq.s (\ref{5.26}) and (\ref{5.27}) one verifies that relation (\ref{6.1}) is obeyed.  
Function $\ddcf$ must also obey the following sum rules
\begin{eqnarray}
\int_0^{\infty}\gamma\pp(r)dr & =& -\gamma\p(0)\,=\,S/4V,\label{6.2}\\
\int_0^{\infty}r\gamma\pp(r)dr & =& \gamma(0)\,=\,1,\label{6.3}\\
(\pi/3)\int_0^{\infty}r^4\gamma\pp(r)dr & =& 4\pi\int_0^{\infty}r^2\gamma\pp(r)dr \,=\,V,\label{6.4}\\
(2\pi\big/15)\int_0^{\infty}r^6\gamma\pp(r)dr & =& 4\pi\int_0^{\infty}r^4\gamma\pp(r)dr \,=\,2\, {R_G}^2. \label{6.5}\end{eqnarray}
The first originates from Porod's law\refup{9}, the second and third from  definition (\ref{1.1}) of
$\gamma(r)$ and the fourth is related to Guinier's law\refup{2} since $R_G$ denotes Guinier's giration radius that, in the octahedron case, is equal to $1\big/5\sqrt{2}$. It has been found that the numerical evaluation
of the integrals, reported on the left hand sides of (\ref{6.2})-(\ref{6.5}), differs, in absolute value, from the right hand side values less than $8 \times 10^{(-15)}$.\\
Using (\ref{6.2}) one finds that the chord length probability density  of the octahedron is
\begeq\label{6.3a}
\cld\, = \, (4V/S)\ddcf.
\endeq
The explicit knowledge of $\ddcf$ allows one to determine numerically both $\gamma\p(r)$ and
$\gamma\p(r)$ by the relations
 \begin{eqnarray}
 \gamma\p(r)&=&-\int_r^{\sqrt{2}}\gamma\pp(t)dt \nonumber\\
\gamma(r)&=&\int_r^{\sqrt{2}}(t-r)\gamma\pp(t)dt.\nonumber
\end{eqnarray}
\begin{figure}[!h]
\includegraphics[width=7.truecm]{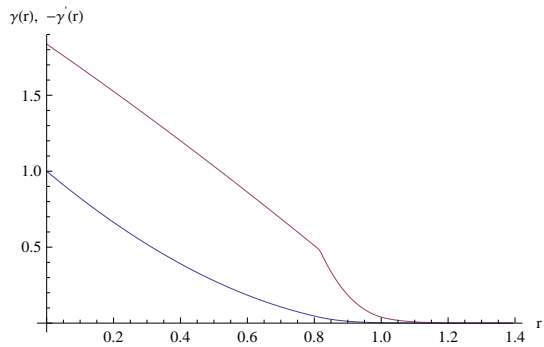}
\caption{\label{Fig9}{{The bottom blue curve is the plot of $\gamma(r)$ and the top red one
that of -$\gamma'(r)$. }}}
\end{figure}
Figure \ref{Fig9} shows the resulting plots of $\gamma(r)$ and  -$\gamma'(r)$. One observes the
well known phenomenon that the particles features become gradually less evident in passing from
$\ddcf$  to $\gamma(r)$.
\section{Conclusion}
It has been show that the chord-length probability density of the regular octahedron can
algebraically be expressed in terms of  elementary functions by separating it into
the contributions due to the pairs of the octahedron facets that share a side, a vertex or
are each other parallel. The final expressions are somewhat longer than  the cube\refup{14} and
the tetrahedron ones\refup{18}, \ie\ the other two Platonic solids for which the $\ddcf$s  have
explicitly been worked out. At this point it is not unreasonable to conjecture that the $\ddcf$s of
all Platonic solids have an algebraic form   and, recalling the parallelepiped result\refup{16}
as well as the result shown in the two-dimensional case for polygons\refup{20},
the conjecture holds likely true for all polyhedrons.\\
\section*{Acknowledgment}
I thank dr. Wilfried Gille for his critical reading of the ms and for having spotted some misprints.  
\vfill\eject
\section*{References}
\begin{description}
\item[\refup{\ 1}]  P. Debye and  A.M. Bueche, {\em J. Appl. Phys.} {\bf 20}, 518, (1949).
\item[\refup{\ 2}] A. Guinier and G.  Fournet,  {\em Small-Angle
Scattering of X-rays},  (Wiley, New York, 1955).
\item[\refup{\ 3}]  L.A. Feigin and  D.I. Svergun,  {\em Structure Analysis
by Small-Angle X-Ray and Neutron Scattering}, (Plenum Press, New York, 1987).
\item[\refup{\ 4}] P. Debye, H.R. Anderson and H. Brumberger,
{\em J. Appl. Phys.},  {\bf 20}, 518,  {(1957)}.
\item[\refup{\ 5}] S. N. Chiu, D. Stoyan,  W.S. Kendall and J. Mecke, {\em Stochastic Geometry and its Applications}, (Wiley, Chichester, 3rd Ed. 2013).
\item[\refup{\ 6}] L.A. Santal\'o,{\em Integral Geometry and Geometric Probability}, (Addison-Wesley, Reading (MA), 1970).
\item[\refup{\ 7}] S. Ciccariello, G. Cocco, A. Benedetti and S. Enzo, {\em Phys. Rev. B},
{\bf 23}, 6474, (1981).
\item[\refup{\ 8}] S. Ciccariello,   {\em J. Math. Phys.} {\bf 36}, 509, (1995).
\item[\refup{\ 9}] G. Porod, {\em Kolloid Z.} {\bf  124}, 83, (1951).
\item[\refup{10}] R. Kirste and G. Porod,   {\em Kolloid Z.} {\bf 184}, 1, (1962).
\item[\refup{11}] H. Wu  and  P. W. Schmidt,  {\em J. Appl. Crystall.} {\bf 7}, 131, (1974).
\item[\refup{12}] S. Ciccariello  {\em Acta Crystall. A} {\bf 41}, 560, (1985).
\item[\refup{13}] S. Ciccariello, {\em Phys. Rev. A}, {\bf 44}, 2975, (1991).
\item[\refup{14}] J.  Goodisman,  {\em J. Appl. Crystall.} {\bf 13}, 132, (1980).
\item[\refup{15}] W. Gille,   {\em Exp. Tech. Phys.} {\bf 35}, 93, (1987).
\item[\refup{16}] W. Gille, {\em J. Appl. Crystall.}  {\bf 32}, 1100, (1999).
\item[\refup{17}] C. Burger and W. Ruland, {\em Acta Crystall. A} {\bf 57}, 482, (2001).
\item[\refup{18}] S. Ciccariello  {\em J. Appl. Crystall.} {\bf 38}, 97, (2005).
\item[\refup{19}] S. Ciccariello and R. Sobry, {\em Acta Crystall. A} {\bf 51}, 60, (1995).
\item[\refup{20}] S. Ciccariello, {\em J. Math. Phys.} {\bf 50}, 103527, (2009).
\end{description}
\end{document}